%% file: Trans2010.tex
\documentclass[journal]{IEEEtran}

\usepackage{cite,graphicx,subfigure,amsmath,amsfonts}

\input{MatrixNotation}

\begin{document}

\title{Regularized Sampling of Multiband Signals}

\author{J.    Selva  \thanks{This work  has    been submitted to   the   IEEE for possible
    publication.  Copyright may  be transferred without notice,  after which this  version
    may no  longer  be accessible. The   author is with  the  Dept.   of Physics,  Systems
    Engineering and Signal  Theory (DFISTS), University of  Alicante, P.O.Box 99,  E-03080
    Alicante,  Spain (e-mail: jesus.selva@ua.es).    This work has   been supported by the
    Spanish Ministry of Education  and Science (MEC),  Generalitat Valenciana (GV), and by
    the  University   of    Alicante (UA)    under   the  following   projects/programmes:
    TEC2005-06863-C02-02,  HA2007-075 and  ``Ram\'{o}n  y Cajal''  (MEC); ACOMP07-087  and
    GV07/214 (GV); and GRE074P (UA).  }}

\maketitle

\markboth{}{}

\begin{abstract}

This paper presents a regularized  sampling method for  multiband signals, that makes it
possible to approach the Landau limit, while keeping  the sensitivity to  noise at a low
level.  The   method is based  on   band-limited  windowing, followed   by trigonometric
approximation in  consecutive time intervals.   The key point  is that the trigonometric
approximation ``inherits'' the multiband property, that is,  its coefficients are formed
by bursts of  non-zero elements corresponding  to the multiband  components. It is shown
that this method can be well combined with the  recently proposed synchronous multi-rate
sampling (SMRS) scheme, given that the resulting  linear system is  sparse and formed by
ones and zeroes.  The proposed method allows one to trade  sampling efficiency for noise
sensitivity, and is specially well suited for bounded signals with unbounded energy like
those in communications, navigation, audio systems, etc.  Besides, it is also applicable
to finite energy signals and  periodic band-limited signals (trigonometric polynomials).
The paper includes a subspace method for blindly estimating the support of the multiband
signal  as well  as its  components,   and the  results  are  validated through  several
numerical examples.

\end{abstract}

\section{Introduction}
\label{sec:i}

Sampling is the  operation  that  makes the  discrete   processing of continuous   signals
possible.  The basic tool for this  operation is Shannon's  Sampling Theorem, which states
that a signal can be reconstructed without error from its samples taken at a rate equal to
twice its maximum frequency (Nyquist rate), \cite{Jerri77}.   In some situations, however,
the signal is multiband, in  the sense that  its spectral support is  composed of a finite
set  of disjoint  intervals.   In this case,  sampling  at the Nyquist   rate can be  very
inefficient.  The design of alternative sampling schemes for this kind of signals has been
an active research topic  for decades.  It was   first demonstrated by   H.  J. Landau  in
\cite{Landau67} that if the signal is sampled nonuniformly, then the average sampling rate
is lower bounded by the measure of its spectral support,  (Landau lower bound).  Later, it
was shown in \cite{Herley99}  that there exist nonuniform  sampling patterns that approach
this bound as  much as desired. The  sampling scheme in  this reference was  the so-called
``multi-coset  sampling'', which consists  in selecting a sampling  rate above the Nyquist
rate, and then choosing a periodic nonuniform subsequence of it.
The subsequent investigations on this topic have mainly focused on devising low complexity
and   stable       (low   sensitivity)   implementations     of     multi-coset  sampling,
\cite{Feng96,Feng96b,Venkataramani00,Venkataramani01}.  
Recently,  compressed sensing techniques   have been applied  to  multi-coset sampling  in
\cite{Mishali08,Mishali09}, so as to detect the band positions, as  well as to reconstruct
each of the  components.  An alternative  method to discretize   multiband signals is  the
random  demodulator in \cite{Tropp10,Mishali10},   in which the   complexity of the analog
processing  is increased, with  the aim of employing  only analog-to-digital converters of
moderate input bandwidth.
Finally, it  is  worth mentioning  the synchronous  multi-rate  sampling (SMRS)  scheme in
\cite{Rosenthal08}.  Here, the multiband  signal is sampled in  several uniform grids with
different rates, in order to produce a bunch of lowpass signals, in  which the spectrum of
the multiband signal appears aliased.  The  key idea is that  this aliasing may not affect
all frequencies of  the original spectral  support for all sampling  rates.  This fact  is
exploited in \cite{Rosenthal08} by properly selecting  the number of sampling frequencies,
so as to  achieve the reconstruction  of the multiband   signal.  This technique  has been
further developed  in a second  paper, which also includes  the application  of compressed
sensing \cite{Fleyer10}.

An assumption that is common to these approaches  is that of  finite energy: the multiband
components  are assumed to be  in the Lebesgue  space $L^2$, and  the machinery of Fourier
analysis   for this space  is  applied extensively.   This way  to  proceed simplifies the
analysis, because signals  in $L^2$ have  a proper spectrum  function, and  the well-known
properties of the Fourier transform apply.  Nevertheless, there is a mismatch between this
model and  the situation encountered  in  multitude applications,  (communications, audio,
navigation, etc).  Quite often, it is not possible to sample a time interval that contains
most of the   signal's energy due   to its long  duration.   This limitation  has negative
effects, that can  be readily noticed  in the single-band case  by analyzing the  Sampling
Theorem.    Given a    finite    energy  signal $\Fs(t)$       whose  spectrum lies     in
$[-1/(2\tau),1/(2\tau)]$, the Sampling Theorem provides the representation
\be{eq:1}
\Fs(n\tau+u)=\sum_{p=-\infty}^{\infty}\Fs((n-p)\tau)\sinc(p+u/\tau),
\ee
where $n$ is an integer and $u$  is any time shift  following $-\tau/2\leq u<\tau/2$. Yet
in practice, the infinite set of samples $\Fs((n-p)\tau)$ is not available, and then
(\ref{eq:1}) must be truncated at some index $P$ 
\be{eq:2}
\Fs(n\tau+u)\approx\sum_{p=-P}^{P}\Fs((n-p)\tau)\sinc(p+u/\tau).
\ee
This last formula is well known for its poor accuracy,  \cite{Butzer82}.  The problem lies
in that (\ref{eq:1}) converge only because the  samples $\Fs((n-p)\tau)$ converge to zero
as $|p|\rightarrow\infty$, due  to the fact  that $\Fs(t)$ has  finite energy.  So, if the
energy of  $\Fs(t)$ is mostly  outside the sampling  interval $[-PT,PT]$, the  accuracy is
poor due to the sinc tails corresponding to the neglected samples.

This problem  is  well known  in the  signal  processing field,  and  it can be  solved by
\emph{regularizing} the interpolation formula in (\ref{eq:2}), using various filter design
techniques,  \cite{Laakso96}.  A typical method   consists in assuming  that  there is some
excess sampling bandwidth,  i.e, that the  spectrum of $\Fs(t)$  lies in $[-B/2,B/2]$ with
$B\tau<1$ (and not $B\tau\leq 1$), and  then multiplying the  sinc samples in (\ref{eq:2})
by a set of weights $\Fv_p(u)$,
\be{eq:316}
\Fs(n\tau+u)\approx\sum_{p=-P}^{P}\Fs((n-p)\tau)\Fv_p(u)\sinc(p+u/\tau).
\ee
This   regularization  converts  the   ill-behaved  interpolator in   (\ref{eq:2})  into a
well-behaved one, and   this  can be  readily   seen in the     error trends: while    the
interpolation error of  (\ref{eq:2}) decreases only as   $1/P$, \cite{Butzer82}, there  are
sets of coefficients $\Fv_p(u)$ for which the error of (\ref{eq:316}) decreases exponentially
as $\Fe^{-\pi (1-B\tau)P}$, \cite{Knab79}.  The price paid for  the regularization is that
it is  not possible to  represent signals using the  full sampling bandwidth  anymore, and
 it is also  not  possible to implement  filters with sharp   transitions, since this would
produce convergence problems similar to those of the sinc series.

By analogy with  the single band  case, the multiband problem  is  currently in the  stage
equivalent  to that  of the  sinc series  in (\ref{eq:1})  and  its  truncated version  in
(\ref{eq:2});  i.e, some sampling  schemes have been identified   that approach the Landau
rate, and allow the recovery of the multiband  signal and its  components, but there is no
regularization procedure  available.    This lack of   regularization  can be   noticed in
analytical devices, like the slicing of  the spectrum using  sharp transitions, and in the
use of discrete sequences with full spectrum, like coset sequences.

The purpose of   this paper is   to present a regularized   sampling scheme for  multiband
signals.  Relative  to  the existing methods   in the literature,  the  one in this  paper
provides   a   flexible interface   between    the multiband   signal  and   its  discrete
representation, that allows one to trade sampling efficiency for noise sensitivity. In the
scheme,  the reconstruction procedure is  specified in the  form of weighted trigonometric
polynomials, which are  able to interpolate  the multiband  signal  and its  components in
consecutive time intervals.

The starting point  in the derivation  of  the scheme  is  the boundness assumption  which
substitutes the usual finite energy assumption, i.e, the  multiband components are assumed
to have bounded time amplitude, but their energy can be  unlimited.  To make no assumption
about the  energy is  meaningful since,  for long  signals,  it usually  lies outside  the
sampling interval and can be arbitrarily  large.  And to  work assuming bounded components
reflects clear physical and  engineering constraints: signals have  bounded peak power due
to restrictions like the maximum transmitted power, or due  to systems like automatic gain
controls.  With the  new assumption, the multiband   components have a  lowpass equivalent
whose real and imaginary parts belong to the Bernstein space $\mathcal{B}_{\pi B}^\infty$,
\cite[chapter  6]{Higgins96}, i.    e, they  can be   described  as  entire  functions  of
exponential type that are real and bounded on the real axis.  This new description has the
drawback  that  the  basic tool   in  spectral analysis,   namely   the Fourier  transform
(Plancherel  theorem), is  not directly  usable.   So, basic  analytic  procedures in  the
literature                          on                   multiband                sampling
\cite{Herley99,Rosenthal08,Fleyer10,Chen10,Tropp10,Mishali10}, like  the division  of  the
spectrum in disjoint intervals,  the use of discrete  sequences  with full spectrum  (like
coset sequences), and the use of  lowpass filters with sharp  transitions are not valid on
bounded band-limited  signals.  However, these  tools can be substituted  by a  simple yet
powerful analytical device, as shown in this paper.  If $\Fz(t)$  is the multiband signal,
the device  consists  in multiplying  $\Fz(t)$  by a   bandlimited  window $\Fw(t)$,  that
approximately ``selects'' a given finite interval, in the sense that its $L^1$ time-domain
content is concentrated in it.  The key result  is that there  exist windows $\Fw(t)$, for
which  the product $\Fz(t)\Fw(t)$ is  a trigonometric polynomial  with negligible error in
the interval selected by $\Fw(t)$.  Besides, the trigonometric polynomial ``inherits'' the
multiband  structure of $\Fz(t)$,  i.e, its  coefficients are  formed by  bursts of values
corresponding    to the bands   of  $\Fz(t)$.  Sec.  \ref{sec:ads}    is dedicated to this
analytical device.

After this, the design of a sampling scheme is addressed in  Sec.  \ref{sec:lss}, where it
is designed for sparse trigonometric polynomials, since the  product $\Fz(t)\Fw(t)$ can be
viewed as a polynomial  of this type with  negligible error. In this  section, it is shown
that a finite  version of the  SMRS scheme produces  a sparse linear  system, in which the
sensitivity to noise can be reduced by increasing the  number of samples.  Afterward, Sec.
\ref{sec:ibm} shows how the finite  sampling grid for  $\Fz(t)\Fw(t)$ can exactly match an
infinite SMRS scheme for $\Fz(t)$.  This makes it possible to interpolate $\Fz(t)$ and its
components  at any $t$.  The interpolation error  is then analyzed  in Sec.  \ref{sec:ea}.
Finally, the blind  sampling problem is  studied in Sec.   \ref{sec:bes}, where the  MUSIC
algorithm from direction of arrival estimation is adopted.

Since the notation   in the paper  is  extensive, the  basic   symbols and operators   are
described in the next sub-section, and there is a list of symbols at the  end of the paper
in Table \ref{tab:1}.
The next section  sketches  the sampling  method,  in order  to give  a  broad view  of the
concepts  involved. The novel  aspects of the  paper  are, fundamentally,  the contents of
Secs. \ref{sec:ads} to \ref{sec:ibm} and Sec. \ref{sec:bes}.

\subsection{Notation}

Several conventions have been adopted in the paper to simplify the notation:

\begin{itemize}

\item Definitions are performed using the operator ``$\equiv$''. 
\item $R(\tau,T)$ denotes the interval
\be{eq:4}
R(\tau,T)\equiv [\tau-T/2,\tau+T/2[
\ee
for arbitrary $\tau$ and $T>0$.

\item The  symbols  $J$, $J'$, $I_{zw}$  and  $I_{sw,m}$ denote  sets  of distinct indices
  (integers).  In $I_{zw}$ and $I_{sw,m}$  the subscripts '$zw$'  and '$sw$' remind one of
  the meaning of these sets: thus, $I_{zw}$ and $I_{sw,m}$ are the index sets of $\Fz(t)$
  and $\Fs_m(t)$, respectively, after windowing using $\Fw(t)$. 

\item $|J|$ denotes the number of elements of a set $J$.

\item The operator ``$\tilde{\;\;\;}$'' denotes periodization with period $T$, 
that is, for a signal $\Fs(t)$, it is 
\be{eq:5}
\tilde{\Fs}(t)\equiv \sum_{p=-\infty}^\infty \Fs(t+pT).
\ee
\item Vectors   and matrices are   written in bold  font  and  in  lower  and  upper case,
  respectively, ($\vm$, $\mm$).

\item $\mi$ denotes the identity matrix. 
 
\item $[\ma]_{h,n}$ denotes the $h,n$ element of matrix  $\ma$; ${[}\vpr]_{r}$ denotes the
  $r$-th element of  vector $\vpr$; and  $[\ma]_{\cdot,n}$ the $n$-th  column of a  matrix
  $\ma$.
\end{itemize}

\section{Sketch of the proposed method}
\label{sec:spm}

Consider a  multiband   signal $\Fz(t)$ formed   by  $M$ bounded  band-limited  components
$\Fs_m(t)$,
\be{eq:6}
\Fz(t)=\sum_{m=1}^M \Fs_m(t),
\ee
where  the band of  $\Fs_m(t)$ is $[a_m,b_m]$,  and  these bands  are known  and appear in
increasing order,  $b_m<a_{m+1}$,   $m=1,\ldots,  M-1$.  The  $m$-th component   can be
viewed as a signal whose baseband version
\be{eq:7}
\Fs_m(t)\Fe^{-j \pi (a_m+b_m)t}
\ee
has real and imaginary parts in the space $\mathcal{B}_{\pi (b_m-a_m)}^\infty$.
The spectral support of $\Fz(t)$ is the set
\be{eq:8}
\mathcal{S}_z\equiv\bigcup_{m=1}^M [a_m,b_m].
\ee
The multiband  sampling problem can be  posed in a finite   time interval, centered  at an
arbitrary instant   $\tau$, $R(\tau,T)$, $T>0$,  in  which a  set  of samples is  taken at
distinct instants $\tau+t_1, \tau+t_2,\ldots, \tau+t_N$,  but assuming that the components
$\Fs_m(t)$ must be  interpolated only inside a   shorter interval $R(\tau,T_1)$,  $T_1<T$.
The objective is then to find a set of functions $\Fh_{m,n}(t)$,  such that the components
$\Fs_m(t)$ and $\Fz(t)$ can be interpolated using
\be{eq:9}
\Fs_m(\tau+t)\approx\sum_{n=1}^N \Fz(\tau+t_n)\Fh_{m,n}(t)
\ee
and
\be{eq:10}
\Fz(\tau+t)\approx\sum_{m=1}^M\sum_{n=1}^N \Fz(\tau+t_n)\Fh_{m,n}(t),
\ee
for $t$ in $R(0,T_1)$.  If there  is a method to  solve this last  problem, then it can be
applied repeatedly  in overlapping intervals  $R(\tau+kT_1,T)$ with integer  $k$, so as to
interpolate $\Fz(t)$ and the components $\Fs_m(t)$ at any $t$.  So, for example, the value
of  any of the components  $\Fs_m(t)$  in a  regular  grid of  instants  $n\Delta t$, with
integer $n$ and arbitrary $\Delta t>0$, could be  obtained by evaluating (\ref{eq:9}) in
the  intersection  of this  grid  with $R(\tau+kT_1,T)$,  and  repeating the procedure for
consecutive values of $k$.

In   this setting, consider   a   band-limited window $\Fw(t)$     with spectrum lying  in
$[-B_w/2,B_w/2]$, $B_w>0$, that  is  \emph{approximately}  time  limited to  the  interval
$R(0,T)$.  Besides,  assume  that it is  bounded
between one    and a  minimum     value greater  than   zero   in  $R(0,T_1)$,   that  is,
$\delta_w\leq\Fw(t)\leq  1$ in $R(0,T_1)$,  $\delta_w>0$.  Note that   $\Fw(t)$ is not the
typical window in filter design, that sharply selects a given time  interval and nulls the
rest of the time axis.   Actually, it is  the dual case:  the spectrum of $\Fw(t)$ sharply
selects  the  band $[-B_w/2,B_w/2]$,  and $\Fw(t)$  is very  small  but  not strictly zero
outside $R(0,T)$.
If $B_w$ is smaller than the minimum separation among bands of $\Fz(t)$, 
\be{eq:11}
B_w<\min_m a_{m+1}-b_m,\;\;m=1,\ldots,M-1,
\ee
then  the  product  $\Fz(\tau+t)\Fw(t)$  is  also  a   multiband  signal  with  components
$\Fs_m(\tau+t)\Fw(t)$,
\be{eq:12}
\Fz(\tau+t)\Fw(t)=\sum_{m=1}^M \Fs_m(\tau+t)\Fw(t),
\ee
and spectral support
\be{eq:13}
\mathcal{S}_{zw}\equiv\bigcup_{m=1}^M [a_m-B_w/2,b_m+B_w/2].
\ee
Thus, $\Fz(\tau+t)\Fw(t)$ is another multiband signal in which the  bands of $\Fz(t)$ have
been expanded, but are still disjoint due to (\ref{eq:11}).   Now, the multiband sampling
problem can  be posed in  (\ref{eq:12}) with  sampling instants  $t_n$  and sample values
$\Fz(\tau+t_n)\Fw(t_n)$.  If there is a satisfactory solution for  this last problem, then
the  original  multiband signal  $\Fz(t)$  and its  components  $\Fs_m(t)$ can  be readily
interpolated in $R(\tau,T_1)$, simply by dividing by $\Fw(t)$.

In the windowed  model in (\ref{eq:12}), the  signal $\Fz(\tau+t)\Fw(t)$  is approximately
time-limited.  This allows  one  to apply  the  Sampling Theorem, but   with the time  and
frequency domains switched, i.e, $\Fz(\tau+t)\Fw(t)$ is  repeated periodically with period
$T$ in the time domain, and this operation \emph{samples} the frequency domain.  Here, the
window  $\Fw(t)$ acts as a  time-domain ``lowpass filter''  with nonuniform response.  The
result of this operation is a trigonometric polynomial, with coefficients equal to samples
of the spectrum of $\Fz(\tau+t)\Fw(t)$,  taken at the frequencies  $p/T$ with integer $p$,
that   lie in  (\ref{eq:13}).  If   $I_{sw,m}$ denotes  the   integer indices  $p$  of the
frequencies $p/T$ inside the band of $\Fs_m(\tau+t)\Fw(t)$,
\be{eq:14}
I_{sw,m}\equiv\{p: a_m-B_w/2\leq p/T\leq b_m+B_w/2,
 p\in\mathbb{Z}\},
\ee
and $I_{zw}$ denotes the union of these sets,
\be{eq:15}
I_{zw}\equiv\bigcup_{m=1}^M I_{sw,m},
\ee
then this approximate sampling theorem says that $\Fz(\tau+t)\Fw(t)$ is the polynomial
\be{eq:16}
\Fz(\tau+t)\Fw(t)\approx \sum_{p\in I_{zw}}\Fc_p(\tau)\Fe^{j2\pi pt/T},
\ee
for $t$ in $R(0,T)$,  where the $\Fc_p(\tau)$  are unknown samples   of the spectrum  of
$\Fz(\tau+t)\Fw(t)$. Besides, the components $\Fs_m(t)$ can be interpolated using
\be{eq:17}
\Fs_m(\tau+t)\approx \frac{1}{\Fw(t)}\sum_{p\in I_{sw,m}}\Fc_p(\tau)\Fe^{j2\pi pt/T},
\ee
for  $t$ in  $R(0,T_1)$.   

\begin{figure}[t]
\begin{flushright}
\includegraphics{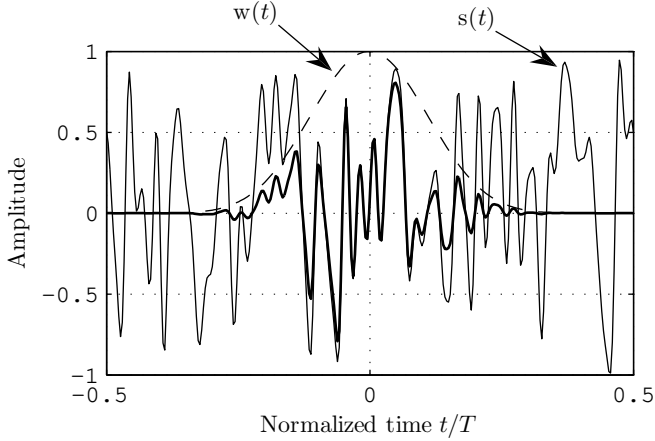}
\end{flushright}
\caption{\label{fig:1}BPSK signal (bandwidth $136/T$, roll off $0.8$, chip rate $0.01324T$), 
window $\Fw(t)$ ($B_wT=13.6$), and windowed signal (bold curve) in
  Sec. \ref{sec:vim}.} 
\end{figure}

\begin{figure}[t]
\begin{flushright}
\includegraphics{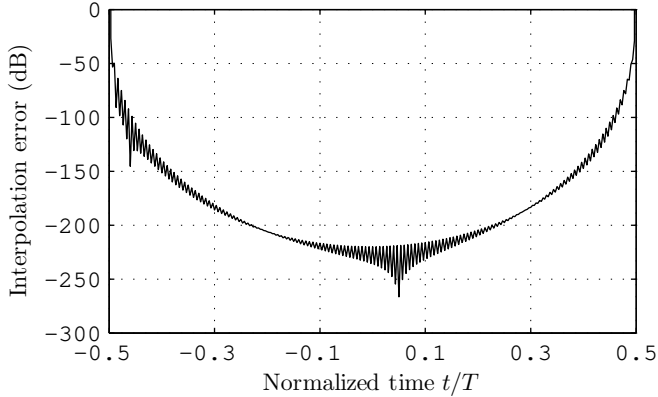}
\end{flushright}
\caption{\label{fig:2} Interpolation  error in validation of  the interpolation  method in
  Sec. \ref{sec:vim}.}
\end{figure}

As will be shown in this paper, the accuracy  of this approximation  depends on the window
$\Fw(t)$. Actually,   it is shown  in  the next  section that  the interpolation  error is
bounded by $A_z\epsilon/\delta_w$, where $A_z$ is a bound on  the amplitude of $\Fz(t)$, and
$\epsilon$  a bound on the summation
\be{eq:18}
\sum_{|p|\neq 0} |\Fw(t+pT)|\leq \epsilon,\;\; t\;\mathrm{in}\; R(0,T).
\ee
The fact is that  there exist  windows,  as the  one presented  in  this paper, for  which
$\delta_w$ grows with $T$, and $\epsilon$  decreases exponentially as $T\rightarrow\infty$
with trend roughly  equal to $\Fe^{-\pi B_w  T}$.  Thus, there is  a moderate value of $T$
for fixed $B_w$  such that (\ref{eq:17})  can be regarded as  an equality, for any  fixed
numerical  precision.   There  are  several  methods  to  construct this   kind of window,
\cite{Helms62,Selva06,Knab79}.  The one  proposed in this paper  in Sec.  \ref{sec:swf} is
based on the  properties of the Fourier  transform of  the  Kaiser-Bessel function.   Fig.
\ref{fig:1} shows this kind of interpolation for a baseband BPSK signal $\Fs(t)$ using the
window  in this  paper.   The bold  curve is  the  windowed signal,  that  is selected for
$\epsilon=10^{-8}$ in  (\ref{eq:18}).  Fig.  \ref{fig:2}  shows  the interpolation error,
which is below -200 dB roughly in $R(0,T/2)$.  (For details, see Sec.  \ref{sec:vim}.)


Coming back to  the formula in  (\ref{eq:16}), if $\epsilon$  is sufficiently small, then
$\Fz(\tau+t)\Fw(t)$ can be  regarded as a trigonometric  polynomial. This implies that the
windowed  problem in Eq.    (\ref{eq:12}) can be   cast in a generic   setting, as that of
computing the coefficients $\beta_p$ of a sparse trigonometric polynomial from its samples
at the abscissas  $t_1, t_2, \ldots,  t_N$.  If $J$   is an index   set, then this kind  of
polynomial can be expressed as
\be{eq:19}
\Falp(t;J)\equiv\sum_{p\in J} \beta_p \Fe^{j2\pi p t/T}.
\ee
The polynomial  on  the  right  in  (\ref{eq:16}) has   this  form with   $J=I_{zw}$  and
$\beta_p=\Fc_p(\tau)$. In this generic setting, the sampling problem  reduces to solving the
linear system
\be{eq:20}
\Falp(t_n;J)\equiv\sum_{p\in J} \beta_p \Fe^{j2\pi p t_n/T},\;\; n=1,\ldots, N.
\ee
If this system has full column rank, then the coefficients $\beta_p$ can be obtained using
\be{eq:21}
\beta_p=\sum_{n=1}^N \eta_{p,n}\Falp(t_n;J),
\ee
where $\eta_{p,n}$ denotes  the pseudo-inverse of (\ref{eq:20}).   Then, the solution  of
the sampling problem for $\Falp(t;J)$ is obtained by  substituting this last equation into
(\ref{eq:19}). This yields
\be{eq:22}
\Falp(t;J)=\sum_{n=1}^N\Falp(t_n;J)\Fthe_n(t;J)
\ee
where 
\be{eq:23}
\Fthe_n(t;J)\equiv\sum_{p\in J} \eta_{p,n}\Fe^{j2\pi p t/T}.
\ee
Besides, since any ``component'' of $\Falp(t;J)$ is specified by a subset $J'$, $J'\subset
J$, its associated   sampling formula is  obtained  by replacing $\Fthe_n(t;J)$   with
$\Fthe_n(t;J')$ in (\ref{eq:22}),
\be{eq:24}
\Falp(t;J')=\sum_{n=1}^N\Falp(t_n;J)\Fthe_n(t;J').
\ee
Finally, the solution  for a sparse  polynomial in (\ref{eq:22})  can  be applied  to the
windowed model in (\ref{eq:12}), if $J'$ is identified with each  of the sets $I_{sw,m}$,
and  $J$ with the  full set of  spectral samples $I_{zw}$.   So, if (\ref{eq:22}) is used
with sample values $\Fz(\tau+t_n)\Fw(t_n)$  and index set $I_{sw,m}$,  and then the effect
of  the window in  (\ref{eq:12}) is removed  by  dividing by  $\Fw(t)$, the  result is an
interpolation formula for $\Fs_m(t)$
\be{eq:25}
\Fs_m(\tau+t)\approx\frac{1}{\Fw(t)}\sum_{n=1}^N \Fz(\tau+t_n)\Fw(t_n)
\Fthe_n(t;I_{sw,m}),
\ee
for $t$ in $R(0,T_1)$.   And the same  can be done  for interpolating $\Fz(t)$, but with
index set $I_{zw}$,
\be{eq:26}
\Fz(\tau+t)\approx\frac{1}{\Fw(t)}\sum_{n=1}^N \Fz(\tau+t_n)\Fw(t_n)
\Fthe_n(t;I_{zw}).
\ee
The last two formulas solve the initial problem in (\ref{eq:9}), with functions
$\Fh_{m,n}(t)$ given by 
\be{eq:27}
\Fh_{m,n}(t)=\Fw(t_n)\Fthe_n(t;I_{sw,m})/\Fw(t).
\ee

This is in broad terms the sampling  method proposed in  this paper for multiband signals,
and  the next  three sections  are  dedicated to  clarify  its  fundamental aspects.   The
windowing and spectral sampling are explained in  the next section.   The key points in it
are  the  approximate truncation  of  the multiband signal   using $\Fw(t)$, and  the dual
sampling  theorem,  which is  demonstrated  by means of  the  Poisson's summation formula.
Also, a specific window  $\Fw(t)$ is selected  in  Sub-section \ref{sec:swf}.  Then,  Sec.
\ref{sec:lss} studies  the selection of the  instants $t_n$ so  that the linear  system in
(\ref{eq:20})  has low  sensitivity  to perturbations.   It turns  out  that a finite SMRS
scheme  produces a sparse linear  system of ones  and zeroes, in  which it  is possible to
reduce the  sensitivity by slightly over-sampling.   Next, Sec.  \ref{sec:bes}  shows that
the finite SMRS scheme in Sec.  \ref{sec:lss} can be perfectly integrated into an infinite
SMRS  scheme for the  initial multiband signal.   This means that  the finite scheme in an
interval  $R(\tau,T)$ with arbitrary  $\tau$ employs samples lying  in the infinite scheme
exclusively.

\section{An approximate   dual   sampling theorem:   approximation  by  means   of
  trigonometric polynomials}
\label{sec:ads}

Recall the  multiband signal  $\Fz(t)$ in  (\ref{eq:6}) with  components $\Fs_m(t)$, and
let $A_{s,m}$ denote specific bounds for them,
\be{eq:28}
|\Fs_m(t)|\leq A_{s,m}.
\ee
Also assume that it is  necessary to approximate the  value of $\Fs_m(t)$  for some or all
$m$, $1\leq m\leq M$, in an interval $R(\tau,T_1)$ for a specific $\tau$.  To perform this
approximation, take   a  band-limited window $\Fw(t)$    in $L^1$ with    spectral support
contained in $[-B_w/2,B_w/2]$, that is bracketed in $R(0,T_1)$ as $\delta_w\leq \Fw(t)\leq
1$ for a specific $\delta_w>0$, and that  additionally fulfills the concentration property
in Eq.  (\ref{eq:18}) with $T_1<T$.
Next, form the new signal 
\be{eq:29}
\Fz_w(t;\tau)\equiv\Fz(\tau+t)\Fw(t).
\ee
This product expands each of the bands $[a_m,b_m]$ of $\Fz(t+\tau)$,  so that the spectral
support of $\Fz_w(t;\tau)$ is contained in the set (\ref{eq:13}).  Besides, this windowing
affects all components $\Fs_m(t)$ of $\Fz(t)$ equally,  i.e, $\Fz_w(t;\tau)$ also consists
of $M$ components $\Fs_{w,m}(t;\tau)$,
\be{eq:30}
\Fz_w(t;\tau)=\sum_{m=1}^M \Fs_{w,m}(t;\tau),
\ee
where  
\be{eq:31}
\Fs_{w,m}(t;\tau)\equiv\Fs_m(t+\tau)\Fw(t). 
\ee
If $B_w$ is smaller than the minimum separation among bands [Eq. (\ref{eq:11})], it turns
out  that $\Fz_w(t;\tau)$  is  also a  multiband signal  (its bands   are disjoint).   The
inequalities in  (\ref{eq:18})  and (\ref{eq:28})   imply  that $\Fs_{w,m}(t;\tau)$  and
$\Fz_w(t;\tau)$  can be    well approximated  by   their  respective  periodic   versions,
$\tilde{\Fz}_w(t;\tau)$ and $\tilde{\Fs}_{w,m}(t;\tau)$,  defined using (\ref{eq:5}). It
is
\bae{eq:32}{1.5}{l}
\D{|\Fs_{w,m}(t;\tau)-\tilde{\Fs}_{w,m}(t;\tau)|}\\
\D{=|\sum_{p\neq 0} \Fs(\tau+t+pT)\Fw(t+pT)|\leq A_{s,m}}
 \epsilon
\eae
and 
\be{eq:33}
|\Fz_w(t;\tau)-\tilde{\Fz}(t)|\leq \sum_{m=1}^M |\Fs_{w,m}(t;\tau)
-\tilde{\Fs}_{w,m}(t;\tau)|\leq 
\epsilon \sum_{m=1}^M  A_{s,m}.
\ee
Now, the signals $\tilde{\Fs}_{w,m}(t;\tau)$  and $\tilde{\Fz}_w(t;\tau)$ are band-limited
and periodic. So, by the Poisson's summation formula \cite[Sec.  2.3]{Higgins96}, they are
sparse trigonometric polynomials, whose coefficients correspond to  nonzero samples of the
spectra of $\Fs_{w,m}(t;\tau)$ and   $\Fz_w(t;\tau)$ at frequencies   of the form   $p/T$,
respectively.   The  sets of  indices  $p$ of  these  frequencies, denoted  $I_{sw,m}$ and
$I_{zw}$,  have  already  been   defined  in  (\ref{eq:14})  and   (\ref{eq:15}).  Thus,
$\tilde{\Fs}_{w,m}(t;\tau)$ and $\tilde{\Fz}_w(t;\tau)$ can be expressed as
\be{eq:34}
\tilde{\Fs}_{w,m}(t;\tau)=\sum_{p\in I_{sw,m}} \Fc_p(\tau) \Fe^{j2\pi p t/T}
\ee
and 
\be{eq:35}
\tilde{\Fz}_w(t;\tau)=\sum_{p\in I_{zw}} \Fc_p(\tau) \Fe^{j2\pi p t/T},
\ee
where $\Fc_p(\tau)$ is the (unknown)  value of the  spectrum of $\Fz_w(t;\tau)$ at frequency
$p/T$.   So, the disjoint    spectral intervals $[a_m,b_m]$    of $\Fz(t)$ translate  into
disjoint bursts of coefficients in $\tilde{\Fz}_w(t;\tau)$.  Note that $\Fs_{w,m}(t;\tau)$
and $\Fz_w(t;\tau)$ have proper spectrum functions, because  they are band-limited signals
that lie in $L^1$.  This is not true for either $\Fs_m(t)$ or $\Fz(t)$.

Finally, the desired formulas are obtained from (\ref{eq:34}) and (\ref{eq:32}). For $t$
in $R(0,T_1)$, it is
\be{eq:36}
\Fs_m(\tau+t)\approx \frac{1}{\Fw(t)}\tilde{\Fs}_{w,m}(t;\tau)=\frac{1}{\Fw(t)}
\sum_{p\in I_{sw,m}} \Fc_p(\tau) \Fe^{j2\pi p t/T}
\ee
with error bounded by $A_{s,m}\epsilon/\delta_w$, and
\be{eq:37}
\Fz(\tau+t)\approx \frac{1}{\Fw(t)}\tilde{\Fz}_w(t;\tau)=
\frac{1}{\Fw(t)}\sum_{p\in I_{zw}} \Fc_p(\tau) \Fe^{j2\pi p t/T}
\ee
with error bounded by $(\sum_m A_{s,m})\epsilon/\delta_w$.

A specific window $\Fw(t)$ is presented in the next sub-section.

\subsection{Selection of a window function}
\label{sec:swf}

\begin{figure}
\begin{center}
\includegraphics{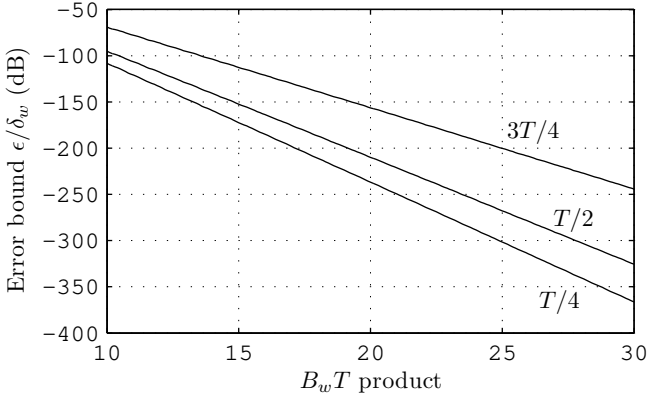}
\end{center}
\caption{\label{fig:10}  Error bound  $\epsilon/\delta_w$ using the  values for $\delta$ and
  $\epsilon$ in (\ref{eq:40}), for $T_1=3T/4, T/2, T/4$.}
\end{figure}

As shown in several   applications \cite{Knab79,Selva08,Selva09,Selva09b}, a  band-limited
window  with excellent  concentration  properties  can be   constructed  from the  Fourier
transform of the Kaiser-Bessel function.  For the interpolation problem in this paper, the
proposed window is
\be{eq:38}
\Fw(t)\equiv \frac{\mathrm{sinc}(\delta B_w t)\mathrm{sinc}\big((1-\delta)B_w 
\sqrt{t^2-\rho^2 T^2/4}\big)}
{\mathrm{sinc}(j(1-\delta)\rho B_wT/2)},
\ee
where
\be{eq:39}
\rho\equiv \sqrt{1-1/(B_w T)^2}.
\ee
In (\ref{eq:38}),  the second sinc  function in the numerator  is the Fourier transform of
the    Kaiser-Bessel  function, and  the   $L^1$  concentration of    $\Fw(t)$  is roughly
proportional to   the  denominator in   (\ref{eq:38}),  which increases   exponentially as
$\Fe^{\pi(1-\delta)\rho B_wT/2}/T$ when $T\rightarrow \infty$.  The first sinc function in
(\ref{eq:38})  is necessary so  as to  damp  the tails  of  $\Fw(t)$, given  that it would
otherwise not belong to $L^1$.  A bound $\epsilon$ for the inequality in (\ref{eq:18}) is
derived in Ap.  \ref{ap:btw} for this window, which is well approximated by
\bae{eq:40}{1.5}{r@{\,\approx\,}l}
\D{\delta_\mathrm{opt}}&\D{0.03326- 0.002084 B_wT + 0.3737\cdot 10^{-4} (B_wT)^2}\\
\D{\epsilon}&\D{10^{1.086- 0.6676 B_wT}.}
\eae
Here, $\delta_\mathrm{opt}$ is the  value of $\delta$ in  (\ref{eq:38}) that achieves  the
corresponding  $\epsilon$ in (\ref{eq:40}).     Note the exponential   rate  in this  last
expression  for  $\epsilon$.  As  a  consequence,  the  model  accuracy  can  be increased
arbitrarily   by increasing $B_wT$.   For  $B_wT=13.61$ it  is  $\epsilon=10^{-8}$ and for
$B_wT=25.59$ it  is $\epsilon=10^{-16}$.  The  actual bound on  the interpolation error is
given  by $\epsilon/\delta_w$, where $\delta_w$  depends on $T_1$.  A value for $\delta_w$
can be obtained numerically using the definition
\be{eq:41}
\delta_w\equiv\inf\limits_{|t|\leq
    T_1/2}\mathrm{w}(t).
\ee
The error  bound $\epsilon/\delta_w$ versus  $B_wT$ is  plotted  in Fig.  \ref{fig:10} for
$T_1=3T/4,  T/2,  T/4$.  In this   figure  $\delta$ and   $\epsilon$  were selected  using
(\ref{eq:40}). Note that any practical accuracy can be achieved  by increasing the product
$B_wT$.

\section{Low sensitivity sampling of a sparse trigonometric polynomial}
\label{sec:lss}

The generic   sampling problem for  sparse  trigonometric polynomials,  as posed   in Sec.
\ref{sec:spm}, Eqs. (\ref{eq:19}) and  (\ref{eq:20}), does not  impose any constraint on
the distribution  of the instants  $t_n$. However, the  condition number of (\ref{eq:20})
can vary wildly with it, as can  be easily checked numerically.  The distribution in which
the sampling points are equally spaced  and cover $R(0,T)$  is interesting, because it
exploits the sparseness  of  the polynomial,  i.e, it  produces  equations in  which  most
coefficients are zero.

To see this point,
assume $\Falp(t;J)$ is uniformly sampled at 
\be{eq:42}
t_{q+1}\equiv t_0+\frac{q}{Q}T,
\ee
for $q=0,1,\ldots, Q-1$,  $Q>0$, where $t_0$  is an arbitrary  instant, and define the
coefficients
\be{eq:43}
\delta_p\equiv \beta_p\Fe^{j2\pi pt_0/T}.
\ee
Next, compute the DFT of the sequence $\Falp(t_{q+1};J)$ (divided by $Q$),  defining a
corresponding set of coefficients $\Lambda_r$:
\bae{eq:44}{1.5}{l}
\D{\Lambda_{r}\equiv\frac{1}{Q}\sum_{q=0}^{Q-1}\Falp(t_{q+1};J)\Fe^{-j2\pi rq/Q}}\\
\D{=\frac{1}{Q}\sum_{q=0}^{Q-1}\sum_{p\in J} \beta_p 
\Fe^{j2\pi pt_{q+1}/T}\Fe^{-j2\pi rq/Q}}\\
\D{=\frac{1}{Q}\sum_{q=0}^{Q-1}\sum_{p\in J} \beta_p 
\Fe^{j2\pi p(t_0/T+q/Q)}\Fe^{-j2\pi rq/Q}}\\
\D{=\sum_{p\in J} \beta_p 
\Fe^{j2\pi pt_0/T}\frac{1}{Q}\sum_{q=0}^{Q-1}\Fe^{j2\pi (p-r)q/Q}}\\
\D{=\sum_{\genfrac{}{}{0pt}{1}{p\in J}{p\equiv r\;(\mathrm{mod}\;Q)}} \beta_p 
\Fe^{j2\pi pt_0/T}=
\sum_{\genfrac{}{}{0pt}{1}{p\in J}{p\equiv r\;(\mathrm{mod}\;Q)}} \delta_p.}
\eae
This derivation shows that $\Lambda_r$ can be computed either from the samples
$\Falp(t_{q+1};J)$ using its definition,
\be{eq:45}
\Lambda_{r}=\frac{1}{Q}\sum_{q=0}^{Q-1}\Falp(t_{q+1};J)\Fe^{-j2\pi rq/Q},
\ee
or from the coefficients $\delta_p$ using 
\be{eq:46}
\Lambda_{r}=\sum_{\genfrac{}{}{0pt}{1}{p\in J}{p\equiv r\;(\mathrm{mod}\;Q)}} 
\delta_p.
\ee
So, if the samples $\Falp(t_{q+1};J)$ are  known, Eq. (\ref{eq:45})  allows one to compute
the  $\Lambda_{r}$  with the best   possible conditioning, given  that  DFT  matrices have
condition number equal to one. And  Eq.  (\ref{eq:46}) shows that  $\Lambda_r$ is just the
sum of some of the coefficients $\delta_p$ (those with index $p$ congruent with $r$ modulo
$Q$).  In some cases, the sum in Eq. (\ref{eq:46}) may contain a  single element, and then
one    of  the   coefficients   $\delta_r$   would  be    revealed   by  $\Lambda_r$, i.e,
$\delta_p=\Lambda_r$ with $p\equiv r  (\mathrm{mod}\;\; Q)$ and   $0\leq r <Q$.   Usually,
however there are several summands, but if the $\Lambda_{r}$ are known, then (\ref{eq:46})
is in any case a sparse linear system of ones and zeroes.

This argument suggests that a regular grid  like (\ref{eq:42}) is  a good choice, in order
to obtain a well conditioned linear system. But the  selection of $Q$  is problematic.  On
the one hand, if $Q$ is larger than or equal to the difference between the largest and the
smallest element in  $J$, then the sum   in (\ref{eq:46}) has   a single summand, and  the
$\Lambda_r$ reveal the coefficients $\delta_p$ completely.  But the over-sampling would be
too large in this case.  On  the other hand,  if $Q$ is  smaller than that difference, the
system in (\ref{eq:46}) could be  under-determined.  The solution  proposed in the sequel
consists in adopting  a finite version  of the SMRS   scheme in \cite{Fleyer10},  in which
several integer moduli $Q_1, Q_2, \ldots, Q_K$ are used instead of a single one.

Let  us state first  the solution of  the interpolation problem  for this scheme, and then
study its sensitivity. For a fixed set of moduli $Q_k$,  let $t_{k,q+1}$ denote a finite
SMRS scheme with initial instant $t_0$,
\be{eq:47}
t_{k,q+1}=t_0+Tq/Q_k,\;\;q=0,1, \ldots, Q_k-1,
\ee
and let $\Lambda_{k,r}$ denote the scaled DFT for modulo $Q_k$ as in (\ref{eq:44}),
\be{eq:48}
\Lambda_{k,r}\equiv\frac{1}{Q_k}\sum_{q=0}^{Q_k-1}\Falp(t_{k,q+1};J)\Fe^{-j2\pi rq/Q_k}.
\ee
Then, Eq. (\ref{eq:46}) for all moduli $Q_k$ can be written as 
\be{eq:49}
\Lambda_{k,r}=\sum_{\genfrac{}{}{0pt}{1}{p\in J}{p\equiv r\;(\mathrm{mod}\;Q_k)}}
 \delta_p, 
\ee
$k=1, 2,\ldots,   K$, $q=0, 1,   \ldots, Q_k$.  This  is   a linear system   with unknowns
$\delta_p$. If it has full column rank, then its  solution can be written  is terms of its
pseudo-inverse, denoted $\lambda_{p,k,r}$, i.e,
\be{eq:50}
\delta_p = \sum_{k=1}^K \sum_{r=0}^{Q_k-1} \lambda_{p,k,r} \Lambda_{k,r}
\ee
Now, Eqs. (\ref{eq:43}) and (\ref{eq:19}) allow one to write $\Falp(t;J)$ in terms
of the coefficients $\delta_p$,
\be{eq:51}
\Falp(t;J)=\sum_{p\in J} \delta_p\Fe^{j2\pi p (t-t_0)/T}.
\ee
Finally, if Eqs. (\ref{eq:48}) and (\ref{eq:50}) are  substituted into this equation, and
the summations are then reordered, the result is
\be{eq:52}
\Falp(t;J)=\sum_{k=1}^K \sum_{q=0}^{Q_k-1}\Falp(t_{k,q+1};J)\Fthe_{k,q}(t;J,t_0),
\ee
where 
\bae{eq:53}{1.5}{l}
\D{\Fthe_{k,q}(t;J,t_0)}\\
{}\hspace{0.5cm}{}\hfill\D{\equiv \frac{1}{Q_k}\sum_{p\in
  J}\Big(\sum_{r=0}^{Q_k-1}\lambda_{p,k,r}\Fe^{-j2\pi rq/Q_k}\Big) \Fe^{j2\pi p (t-t_0)/T}.}
\eae
The last two equations specify the solution to the interpolation  problem using the finite
SMRS scheme. Note that (\ref{eq:52}) can be used to reconstruct any component of
$\Falp(t;J)$, simply by restricting the index set in $\Fthe_{k,q}(t;J,t_0)$, i.e, if
$J'\subset J$, then
\be{eq:54}
\Falp(t;J')=\sum_{k=1}^K \sum_{q=0}^{Q_k-1}\Falp(t_{k,q+1};J)\Fthe_{k,q}(t;J',t_0).
\ee
Also, note that here $\Fthe_{k,q}(t;J,t_0)$ is the specific reconstruction function in the
SMRS scheme with initial  instant $t_0$, while  $\Fthe_n(t;J)$ in Sec.  \ref{sec:spm}, Eq.
(\ref{eq:22}), is a reconstruction function for a  generic sampling set $t_1, t_2,\ldots,
t_N$ and fixed $\tau$.

The selection of the moduli $Q_k$ can be done now  so as to make  this method viable, i.e,
so that the linear system  in (\ref{eq:49}) has  full column rank,  and the sensitivity to
perturbations of (\ref{eq:52}) is low. Let us define first the sensitivity measure
\be{eq:55}
\Fgam(t;J',t_0)\equiv ( \sum_{k=1}^K \sum_{q=0}^{Q_k-1}|\Fthe_{n}(t_{k,q};J',t_0)|^2)^{1/2}.
\ee
and its supremum in $R(0,T)$,
\be{eq:56}
\Fgam(J')\equiv\sup_{-T/2\leq t<T/2} \Fgam(t;J',0).
\ee
$\Fgam(t;J',t_0)$ is the standard deviation factor due to any perturbations in the samples
$\Falp(t_{k,q};J')$. So,  if these samples  are contaminated by  perturbations of zero mean
and variance $\sigma^2$, then the value of $\Falp(t;J')$ computed using (\ref{eq:54}) has
variance $\sigma^2\Fgam^2(t;J',t_0)$.

The following method gives a suitable collection  of moduli $Q_k$:

\begin{enumerate}
\item Select  positive integers $Q_1$ and  $K$ following $Q_1K\approx  |J|$ and then set
  $Q_k=Q_1+k-1$, $k=2,\ldots,K$.

\item Check whether the linear system
\be{eq:57}
\Lambda_{k,r}=\sum_{\genfrac{}{}{0pt}{1}{p\in J}{p\equiv r\;(\mathrm{mod}\;Q_k)}} \delta_p 
\ee
is over-determined  (or determined).  If  not increase either $K$ or
$Q_1$.

\item Finally, the sensitivity can be reduced by adding other $Q$ which are selected so as
  to minimize $\Fgam(J)$, or $\Fgam(J')$ if the performance for a specific components
  $J'\subset J$ must be improved.
\end{enumerate}

\section{Interpolation of a bounded multiband signal using the SMRS scheme}
\label{sec:ibm}

As shown in Sec.  \ref{sec:ads},  Eq.  (\ref{eq:37}), the  sampling problem for a bounded
multiband signal $\Fz(t)$ can be  reduced to a problem  of the same  type, but involving a
sparse  trigonometric  polynomial. Specifically, it was shown that there is a polynomial
$\tilde{\Fz}(t;\tau)$ such that 
\be{eq:58}
\Fz(\tau+t)\approx \frac{\tilde{\Fz}(t;\tau)}{\Fw(t)},
\ee
where $t$ lies in $R(0,T)$. This  formula specifies how the  multiband signal $\Fz(t)$ can
be interpolated if the polynomial  $\tilde{\Fz}(t;\tau)$ is known,  and how the polynomial
$\tilde{\Fz}(t;\tau)$ can be obtained from samples of $\Fz(t)$.

The  objective  of this  section is  to show   that  the finite SMRS scheme in the
previous section of the form
\be{eq:59}
t_0+Tq/Q_k, \;\;q=0,  1, \ldots, Q_k-1,\;\; k=1, 2,\ldots, K,
\ee
can be integrated into the \emph{infinite} SMRS scheme
\be{eq:60}
nT+Tq/Q_k,\;\; n\;\;\mathrm{in}\;\;\mathbb{Z},\;\;q=0,  1, \ldots, Q_k-1.
\ee
More precisely, there  is a $t_0$  such that there is  a one-to-one correspondence between
the samples of $\Fz(t)$ in the scheme (\ref{eq:60}) that lie in $R(\tau,T)$, and those of
$\tilde{\Fz}(t;\tau)$ in  the scheme (\ref{eq:59}).  To  demonstrate this, assume that it
is necessary to sample $\tilde{\Fz}(t;\tau)$ at one of the instants in (\ref{eq:59}). Let
$n_{k,q}(t_0)$ denote the  unique integer such that   $t_0+Tq/Q_k+ n_{k,q}(t_0)T$ lies  in
$R(0,T)$,
\be{eq:61}
n_{k,q}(t_0)\equiv\lceil (\tau-t_0)/T-1/2-q/Q_k\rceil.
\ee
From (\ref{eq:58}) and recalling that $\tilde{\Fz}(t;\tau)$ is $T$-periodic, it is
\bae{eq:62}{1.5}{l}
\D{\tilde{\Fz}(t_0+Tq/Q_k;\tau)=\tilde{\Fz}(t_0+Tq/Q_k+n_{k,q}(t_0)T;\tau)}\\
\D{\approx \Fz(\tau+t_0+Tq/Q_k+n_{k,q}(t_0)T)}\\
\D{{}\hfill\cdot \Fw(t_0+Tq/Q_k+n_{k,q}(t_0)T)}
\eae
The argument of $\Fz$ in this formula belongs to the scheme in (\ref{eq:60}) for any pair
of indices $(k, q)$ if $t_0=-\tau$. So, for this $t_0$, it is
\bae{eq:63}{1.5}{l}
\D{\tilde{\Fz}(-\tau+Tq/Q_k;\tau)\approx\Fz(Tq/Q_k+n_{k,q}(-\tau)T)\hspace{0.5cm}{}}\\
\D{{}\hfill\cdot \Fw(-\tau+Tq/Q_k+n_{k,q}(-\tau)T).}
\eae
This  is the one-to-one   correspondence between the   samples of $\tilde{\Fz}(t;\tau)$ in
(\ref{eq:59}) and  those    of $\Fz(t)$ in   the   intersection of  (\ref{eq:60})   with
$R(\tau,T)$.

Now, it is possible to write an interpolation formula for $\Fz(t)$. For  this it is enough
to   ``sample''  $\tilde{\Fz}(t)$   using   (\ref{eq:63}),  then  obtain   the  polynomial
$\tilde{\Fz}(t)$ approximately using the formula in (\ref{eq:54}), and finally undoing the
windowing [dividing by $\Fw(t)$]. To simplify the notation, define first the instants
\be{eq:64}
t'_{k,q}(\tau)\equiv -\tau+Tq/Q_k+n_{k,q}(-\tau)T.
\ee
From (\ref{eq:63}), the samples of $\tilde{\Fz}(t;\tau)$ are
\be{eq:65}
\tilde{\Fz}(-\tau+Tq/Q_k;\tau)\approx \Fz(\tau+t'_{k,q}(\tau))\Fw(t'_{k,q}(\tau)).
\ee
If the sampling formula in (\ref{eq:54}) is applied to $\tilde{\Fz}(t;\tau)$ with these
sample values, it is 
\be{eq:66}
\tilde{\Fz}(t;\tau)\approx \sum_{k=1}^K
\sum_{q=0}^{Q_k-1}\Fz(\tau+t'_{k,q}(\tau))\Fw(t'_{k,q}(\tau))
\Fthe_{k,q}(t;I_{zw},-\tau)
\ee
And, finally, (\ref{eq:58}) gives the desired formula for $\Fz(t)$,
\bae{eq:67}{1.5}{l}
\D{\Fz(\tau+t)\approx\frac{1}{\Fw(t)}\sum_{k=1}^K
\sum_{q=0}^{Q_k-1}}\\
\D{{}\hspace{1.5cm}\Fz(\tau+t'_{k,q}(\tau))\Fw(t'_{k,q}(\tau))
\Fthe_{k,q}(t;I_{zw},-\tau).}
\eae
The formula for $\Fs_m(t)$ is the same but with $I_{zw}$ replaced with $I_{sw,m}$. 
  
\section{Error analysis}
\label{sec:ea}

In the error analysis in the sequel, the sampling  scheme is denoted  using a single index
as  $t_1, t_2,\ldots, t_N$  instead of  two, so  as  to simplify  the  notation. This also
affects  the function   $\Fthe_{k,q}(t,J,t_0)$ in  (\ref{eq:53}),   that will be   denoted
$\Fthe_n(t,J,t_0)$.   Also,  the error  terms  will be  written inside   square braces for
readability.

Let us  recall  the sampling and   interpolation process on $\Fz(t)$   in the  last  three
sections, but keeping track  of the different  errors, so as to  produce an error bound.
First, the multiband    signal  $\Fz(t)$ can   be    approximated using  the    polynomial
$\tilde{\Fz}_w(t;\tau)$, [Eq. (\ref{eq:37})],
\be{eq:68}
\Fz(\tau+t)=\frac{1}{\Fw(t)}\tilde{\Fz}_w(t;\tau)+
\Big[\frac{1}{\Fw(t)}(\Fz_w(t;\tau)-\tilde{\Fz}_w(t;\tau))\Big].
\ee
Next, $\tilde{\Fz}_w(t;\tau)$ can be reconstructed using the sampling formula in
(\ref{eq:52}) with $t_0=-\tau$, $J=I_{zw}$, and samples $\tilde{\Fz}_w(t_n;\tau)$,
\be{eq:69}
\tilde{\Fz}_w(t;\tau)=\sum_{n=1}^N \tilde{\Fz}_w(t_n;\tau)\Fthe_n(t;I_{zw},-\tau).
\ee
The samples $\tilde{\Fz}_w(t_n;\tau)$ are approximately equal to $\Fz(\tau+t_n)\Fw(t_n)$:
\be{eq:70}
\tilde{\Fz}_w(t_n;\tau)=\Fz(\tau+t_n)\Fw(t_n)+\Big[\tilde{\Fz}_w(t_n;\tau)-\Fz_w(t_n;\tau)\Big].
\ee
But  $\Fz(\tau+t_n)$ may be contaminated by a perturbation, denoted
$\Feta(t_n)$:
\be{eq:71}
\Fz(\tau+t_n)=\Fz(\tau+t_n)+\Feta(t_n)+\Big[-\Feta(t_n)\Big].
\ee
Next,   replace in  (\ref{eq:70}) the   term  $\Fz(\tau+t_n)$  with   the  right side  of
(\ref{eq:71}),  so as to obtain
\bae{eq:72}{1.5}{l}
\D{\tilde{\Fz}_w(t_n;\tau)=(\Fz(\tau+t_n)+\Feta(t_n))\Fw(t_n)}\\
\D{{}\hspace{1cm}+
\Big[-\Feta(t_n)\Fw(t_n)+\tilde{\Fz}_w(t_n;\tau)-\Fz_w(t_n;\tau)\Big].}
\eae
In turn,  substitute this equation   into (\ref{eq:69}), and   finally use the  result to
replace the   term  $\tilde{\Fz}_w(t;\tau)$ in  (\ref{eq:68}).    The outcome   of  these
replacements is a formula with error term for $\Fz(\tau+t)$:
\bae{eq:73}{1.5}{l}
\D{\Fz(\tau+t)=\frac{1}{\Fw(t)}\sum_{n=1}^N (\Fz(\tau+t_n)+\Feta(t_n))
\Fw(t_n)\Fthe_n(t;I_{zw},-\tau)}\\
\D{+\Bigg[\frac{1}{\Fw(t)}\sum_{n=1}^N  (-\Feta(t_n)\Fw(t_n))
   \Fthe_n(t;I_{zw},-\tau)}\\
\D{+\frac{1}{\Fw(t)}\sum_{n=1}^N  (\tilde{\Fz}_w(t_n;\tau)-\Fz_w(t_n;\tau)) 
\Fthe_n(t;I_{zw},-\tau)}\\
\D{+\frac{1}{\Fw(t)}(\Fz_w(t;\tau)-\tilde{\Fz}_w(t;\tau))\Bigg].}\\
\eae
Now, since  $\Fw(t)\leq 1$ in  $R(0,T_1)$, the three  error  terms can  be bounded  in the
following way. For the first term in (\ref{eq:73}):
\bae{eq:74}{1.5}{l}
\D{
\Big|\frac{1}{\Fw(t)}\sum_{n=1}^N  (-\Feta(t_n)\Fw(t_n))\Fthe_n(t;I_{zw},-\tau)\Big|
\hspace{1cm}{}}\\
\D{{}\hfill\leq 
\frac{A_\eta}{\Fw(t)}
\Fgam(t;I_{z,w},-\tau),}
\eae
where
\be{eq:75}
A_\eta\equiv\Big(\sum_{n=1}^N |\Feta(t_n)|^2\Big)^{1/2}
\ee
and
\be{eq:76}
\Fgam(t;I_{z,w},t_0)\equiv \Big(\sum_{n=1}^N |\Fthe_n(t;I_{zw},t_0)|^2\Big)^{1/2}.
\ee
For the second,
\bae{eq:77}{1.5}{l} \D{\Big|\frac{1}{\Fw(t)}\sum_{n=1}^N
  (\tilde{\Fz}_w(t_n;\tau)-\Fz_w(t_n;\tau))\Fthe_n(t;I_{zw},-\tau)\Big|}\\
\D{{}\hfill\leq\frac{\epsilon\sqrt{N}}{\Fw(t)}\Big(\sum_{m=1}^M
  A_{s,m}\Big)\Fgam(t;I_{z,w},-\tau).  } \eae
And for the third,
\be{eq:78}
\Big|\frac{1}{\Fw(t)}(\Fz_w(t;\tau)-\tilde{\Fz}_w(t;\tau))\Big|\leq \frac{\epsilon}{\Fw(t)}
\sum_{m=1}^M A_{s,m}.
\ee
So, the final bound is 
\bae{eq:79}{1.5}{l}
\D{\Big|\Fz(\tau+t)}\\
\D{{}\hfill-\frac{1}{\Fw(t)}\sum_{n=1}^N (\Fz(\tau+t_n)+\Feta(t_n))
\Fw(t_n)\Fthe_n(t;I_{zw},-\tau)\Big|}\\
\D{\leq \frac{A_\eta \Fgam(t;I_{z,w},-\tau)}{\Fw(t)}}\\
\D{{}\hfill+
\frac{ \epsilon}{\Fw(t)}\Big(\sum_{m=1}^M
  A_{s,m}\Big)\Big(\sqrt{N}\Fgam(t;I_{z,w},-\tau)+1\Big).
}\\
\eae
The first term is due to the  sensitivity of the linear  system. The second  is due to the
mismatch and  perturbations, and it  can be arbitrarily  small, since $\epsilon$  tends to
zero exponentially with $T$, [Eq.  (\ref{eq:40})]. 

The error  analysis  can be  repeated for  the components  $\Fs_m(t)$,  following the same
method, and the result is
\bae{eq:80}{1.5}{l}
\D{\Big|\Fs_m(\tau+t)}\\
\D{{}\hfill-\frac{1}{\Fw(t)}\sum_{n=1}^N (\Fz(\tau+t_n)+\Feta(t_n))
\Fw(t_n)\Fthe_n(t;I_{sw,m},-\tau)\Big|}\\
\D{\leq \frac{A_\eta \Fgam(t;I_{sw,m})}{\Fw(t)}
}\\
\D{{}\hfill +
\frac{ \epsilon}{\Fw(t)}\Big(\Big(\sum_{m=1}^M
  A_{s,m}\Big)\sqrt{N}\Fgam(t;I_{sw,m},-\tau)+A_{s,m}\Big).
}\\
\eae

\section{Blind estimation of the spectral support}
\label{sec:bes}

If the spectral  support of the  multiband signal is  unknown, then it  is not possible to
determine the set $I_{zw}$, and thus  the procedure presented in  the previous sections is
not applicable in  principle.  In the  literature \cite{Mishali08,Fleyer10}, this  problem
has been addressed using  compressed sensing  techniques.  For  the  SMRS scheme in  Secs.
\ref{sec:lss} and \ref{sec:ibm}, this approach would lead to a linear  system like that in
(\ref{eq:57}),  but in which  the  $l^1$ norm of   the set of   coefficients $d_p$ must be
minimized
\be{eq:81}
\{\delta_p\}=\arg \min \sum_{p=P_1}^{P_2} |\delta_p|\;\;\mathrm{subject\; to}\;\;
\Lambda_{k,r}=\sum_{p=P_1}^{P_2} \delta_p. 
\ee
Here, $\{\delta_p\}$ denotes the set of coefficients $\delta_p$, and the frequency indices
are  assumed to  lie  in $[P_1,P_2]$.    However, the  multiband  signal  $\Fz(t)$  can be
approximated in   several intervals  $R(\tau_h,T)$,  $h=1,  2,\ldots,   H$, with  distinct
$\tau_h$ using  polynomials $\tilde{\Fz}_{w}(t;\tau_h)$.  These  polynomials have  the same
index support $I_{zw}$  and, besides, the $\tau_h$  can be selected  in the infinite  SMRS
scheme, so that the relative sampling instants $t_{k,q}$ are the same  for all of them. In
this setting, the blind estimation problem  can be viewed as  a compressed sensing problem
with multiple  measurements  vectors  (MMV),  \cite{Cotter05}.   Nevertheless,  since  the
duration of the multiband signal can be arbitrarily large, the  blind problem can be posed
using any of the techniques from  direction of arrival  (DOA) estimation, \cite[chapters 8
and 9]{VanTreesP4}.  This alternative solution is analyzed in  the next sub-section, where
the MUSIC algorithm is adopted.  It is then tested numerically in Sec.  \ref{sec:bs}.

A final aspect is whether the SMRS scheme  leads to a full  column rank linear system, for
any set   $I_{zw}$.   In the   literature,  a sampling  scheme   of this kind    is termed
``universal'',  \cite{Mishali08,Fleyer10}.  It turns  out that the  finite  SMRS scheme in
Sec. \ref{sec:lss} can be  converted into a  universal one in a  simple way. To see  this,
recall the explicit expression for $\tilde{\Fz}_w(t;\tau)$ in (\ref{eq:35}),
\be{eq:82}
\Fz_w(t;\tau)\approx \tilde{\Fz}_w(t;\tau)=\sum_{p\in I_{zw}} \Fc_p(\tau) \Fe^{j2\pi p t/T},
\ee
and assume that the sampling scheme is multi-coset sampling with repetition period $T$ and
minimum  spacing  $\Delta  t$.  This  means   that $\tilde{\Fz}_w(t;\tau)$  is  sampled at
instants $t_0+n_r \Delta t$, $r=1, 2, \ldots,  R$, where the  $n_r$ are distinct integers,
$R\Delta t=T$, and  $R\geq |I_{zw}|$.  Evaluating  (\ref{eq:82}) at the sampling instants
yields
\be{eq:83}
\tilde{\Fz}_w(t_0+n_r\Delta t;\tau)= \sum_{p\in I_{zw}} \Fc_p(\tau)\Fe^{j2\pi p t_0/T} 
\Fe^{j2\pi p n_r/R}.
\ee
This linear system  may not have  full column rank.   However, note that (\ref{eq:82})  is
also  valid if $T$ is  replaced with a $T'$  following  $T'> T$,  given the  $L^1$ norm of
$\Fz_w(t;\tau)$  in $R(0,T')$ is  larger than in $R(0,T)$,  and the interpolation accuracy
was already  enough in $R(0,T)$.  To  replace $T$ with $T'$  may slightly increase the
number of elements of $|I_{zw}|$, i.e, the spectral sampling will be finer, and in general
there will be two different sets $I_{zw}(T)$ and  $I_{zw}(T')$. However, if the difference
$R-|I_{zw}(T)|\geq  0$  is  large  enough,    then  one  may  expect   that  it  is   also
$R-|I_{zw}(T')|\geq 0$, i.e, the linear  system for the larger  period $T'$ will also have
more equations than unknowns. Knowing this, take $T'=\Delta t P$ where $P$ is the smallest
prime number following $P\geq R$, and evaluate (\ref{eq:82}) with $T'$ in place of $T$,
\be{eq:84}
\Fz_w(t_0+n_r\Delta t;\tau)\approx \sum_{p\in I_{zw}} \Fc_p(\tau) \Fe^{j2\pi p t_0/T'}
\Fe^{j2\pi p n_r/P}.
\ee
If  $R-|I_{zw}(T')|\geq  0$  this  system has   full  column rank,   since all  minors  of
trigonometric Vandermonde matrices of prime order have full rank, by Chebotarev's theorem;
(see \cite[page  25]{Prasolov94}).   So, for $\Fz_w(t;\tau)$   one can turn  a multi-coset
sampling scheme into a universal one by  slightly increasing $T$.   The only condition for
this is  that there must be  a  minimum over-sampling, in   the sense that  the difference
$R-|I_{zw}(T')|$ must not be too small.

Finally, note that the SMRS scheme in Sec.   \ref{sec:lss} is a  multi-coset scheme if one
sets $t_0=-\tau$ and
\be{eq:85}
\Delta t=T/(\nu \,\mathrm{lcm}_k  Q_k), 
\ee
where $\mathrm{lcm}$ denotes the least common multiple and  $\nu$ is any positive integer.
So, the method   of perturbing  $T$  also  works for  SMRS   sampling.  In  practice,  the
difference  between  using   $T$  or  $T'$  can  be  immaterial.   For   example,  in Sec.
\ref{sec:ppn}, the SMR scheme uses the $Q$ values in (\ref{eq:86}) with
\be{eq:87}
\prod_{k=1}^K Q_k=159049016335440. 
\ee
This product is  one of the  possible denominators in (\ref{eq:85})  for a specific $\nu$,
and the smallest prime $P$ following $P\geq \prod_{k=1}^K Q_k$ is
\be{eq:88}
P=159049016335453.
\ee
So, the relative perturbation of $T$ is 
\be{eq:89}
\frac{T'-T}{T}=\frac{P}{R}-1=8.17\cdot 10^{-14}, 
\ee
which is well  below the numerical  precision in that section. 

\subsection{Blind sampling of a collection of sparse trigonometric polynomials with common
index support}
\label{sec:bsc}

Consider  a collection of  trigonometric polynomials  $\Falp_1(t;J), \Falp_2(t;J), \ldots,
\Falp_H(t;J)$ with common index support $J$ and coefficients $\beta_{h,p}$  of the form in
(\ref{eq:19}),
\be{eq:90}
\Falp_h(t;J)\equiv\sum_{p\in J} \beta_{h,p} \Fe^{j2\pi p t/T}.
\ee
Assume that each polynomial is sampled at
the distinct epochs $t_1, t_2, \ldots, t_N$. If the samples are denoted by
$\hat{\alpha}_{h,n}$,
\be{eq:91}
\hat{\alpha}_{h,n}=\alpha_{h}(t_n;J)+\mu_{h,n},
\ee
where $\mu_{h,n}$ is   an  unknown perturbation,   then  Eq.  (\ref{eq:90}) yields    the
following model for $\hat{\alpha}_{h,n}$,
\be{eq:92}
\hat{\alpha}_{h,n}=\sum_{r=1}^{|J|}\beta_{h,p(r)}\Fe^{j2\pi p(r) t_n/T}+\mu_{h,n},
\ee
where $p(r)$ is an ordering of $J$, i.e,  $p(r)$ runs through all  elements of $J$ for
$r=1, 2,\ldots, |J|$. Next, it is straight-forward to convert this equation into a matrix
model, similar to those in DOA estimation. For this, define the matrices
\be{eq:93}
\renewcommand{\arraystretch}{1.5}
\begin{array}{l@{\;}l@{\;}l}
[\ma]_{n,h}\equiv\hat{\alpha}_{h,n},& [\mb]_{r,h}\equiv \beta_{h,p(r)}, & 
[\mn]_{n,h}\equiv\mu_{h,n}, \\
 {[}\vpr]_{r}\equiv p(r),& [\vphi(p)]_n\equiv \Fe^{j2\pi p t_n/T}, &
[\mphi(\vpr)]_{\cdot,r}\equiv \vphi([\vpr]_r).
\end{array}
\ee
The model is 
\be{eq:94}
\ma=\mphi(\vpr)\mb+\mn.
\ee

The purpose now is to estimate the elements of $\vpr$, which  specify the spectral support
of the  polynomials $\tilde{\Fz}_w(t;\tau_h)$, and  the matrix of  coefficients $\mb$.  In
principle, it is possible to apply to  (\ref{eq:94}) a wide  variety of algorithms for DOA
estimation  in  array processing,  \cite[Chapters   8 and  9]{VanTreesP4}.   However, this
estimation/detection problem has a few  differences relative to  the typical DOA problems:
the length of $\vpr$ can be large (on the hundreds), the components of this vector must be
integers  and, as will be  shown in the  numerical examples, the  problem of detecting the
number of components of $\vpr$ is not critical.

An algorithm that seems well suited for (\ref{eq:94}) is MUSIC, \cite{Schmidt86}.  If $P$
denotes an upper bound for $|I_{zw}|$, and $\mur$ is an $N\times (N-P)$ matrix that
spans   the  noise  subspace  of   $\ma$,  $\mur\Herm\mur=\mi$,  then  the
normalized MUSIC spectrum is
\be{eq:95}
\Fchi(p)\equiv \frac{||\vphi(p)||^2}{||\mur\Herm\vphi(p)||^2}.
\ee
The effectiveness of this estimator will be demonstrated in the numerical example in
Sec. \ref{sec:bs}. 

\section{Numerical examples}
\label{sec:ne}

\subsection{Validation of the interpolation method}
\label{sec:vim}

In Secs.  \ref{sec:spm} and \ref{sec:ads}, a basic argument was that a band-limited signal
which is concentrated  in a  time  interval can  be well  approximated by  a trigonometric
polynomial, and this  concentration was achieved  by means of  a window function.  To test
this  property numerically, a BPSK  signal $\Fs(t)$ was  generated, whose modulating pulse
was  a  raised cosine with   roll-off factor 0.8,  peak  amplitude $A_s=1$,  and bandwidth
$B=136/T$, (chip period $T_c=0.01324T$).  The formula for this signal is
\be{eq:317}
\sum_{p=-\infty}^\infty a_p \Fg_{rc}(t-pT_c),
\ee
where $\Fg_{rc}(t)$ is the raised-cosine pulse
\be{eq:318}
\Fg_{rc}(t)\equiv \pi\, \mathrm{\sinc} \Big(\frac{t}{T_c}\Big)
\frac{\cos(\pi\beta t/T_c)}{1-(2\beta t/T_c)^2},
\ee
and the coefficients $a_p$ took two real values  $\pm A'$ at random.  $A'$ was selected so
that the peak amplitude of the signal is  $A_s$.  Then, the window  in Sec.  \ref{sec:swf}
was applied to this signal  with $B_w=13.6/T$ and $\delta=0.0103$  at $\tau=0$. This gives
$\epsilon=10^{-8}$.  Fig. \ref{fig:1} shows the BPSK signal, the  window $\Fw(t)$, and the
windowed signal (bold line). The BPSK signal can be regarded as  a multiband signal with a
single component and   spectral support $[-B/2,B/2]$.     Therefore, the results   in Sec.
\ref{sec:ads} apply, and there is a set of coefficients $\Fc_p(0)$ such that
\be{eq:96}
\Fs(t)\approx \frac{1}{\Fw(t)}\sum_{p=-p_B}^{p_B}\Fc_p(0)\Fe^{j2\pi p t/T}
\ee
where   $p_B\equiv \lfloor (B+B_w)T/2\rfloor$.     In order to  check   whether this is an
accurate  model,  the   signal   $\Fs(t)\Fw(t)$ was    sampled with   rate  $1/(4(B+B_w))$
(over-sampling factor 4), and then  a polynomial like that  on the right of (\ref{eq:96})
was fitted to these data using least squares.  Then, the resulting polynomial, but divided
by $\Fw(t)$, was used to approximate the initial BPSK  signal.  Fig.  \ref{fig:2} presents
the  error of this  procedure.  Note that it  increases toward the  interval limits due to
$\Fw(t)$,  but  the interpolation error  is  very small.   Actually,  for  $t$  in roughly
$R(0,T/2)$, it is below -200 dB.

\subsection{Performance in the presence of noise}
\label{sec:ppn}

\begin{figure}
\begin{center}
\includegraphics{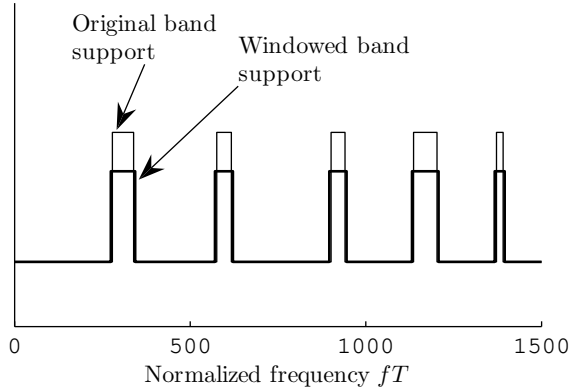}
\end{center}
\caption{\label{fig:3} Supports of the multiband signal before and after windowing in Sec.
  \ref{sec:ppn}. The amplitude along the $y$-axis has no meaning. For the numerical values
  that specify these bands, see table in Sec.  \ref{sec:ppn}.}
\end{figure}

\begin{figure}
\centerline{\includegraphics{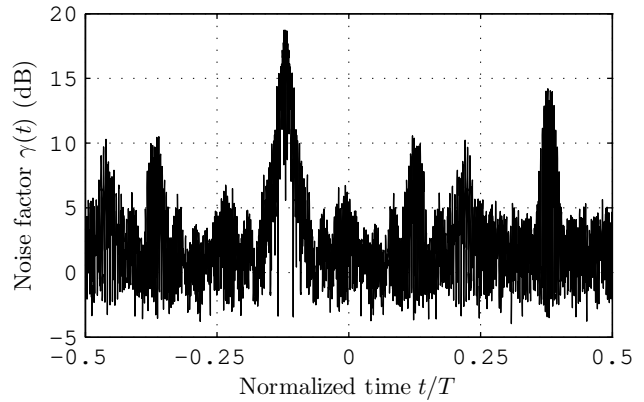}}
\caption{\label{fig:4} Noise factor in example of Sec. \ref{sec:ppn}.}
\end{figure}

In order to test the performance in the presence of noise, a multiband signal with five
components was generated. Its components were signals of one of the following types:
\begin{enumerate}

\item QPSK signal modulated by a raised cosine pulse with roll-off factor 0.8. This signal
  has the form in  Eq. (\ref{eq:317}), but  with symbols in  the constellation $\pm  A'\pm
  jA'$. $A'$ was selected so that the peak absolute value of the signal was one.

\item Sum of 150 undamped exponentials with random phases and frequencies,
\be{eq:319}
\sum_{p=1}^{150} a_p \Fe^{j(2\pi f_p t+\phi_p)}.
\ee
The phases were taken  at random in  the interval $[0,2\pi]$,  and the amplitudes $a_p$ in
the interval $[0,1]$. These amplitudes were then scaled so that the  peak amplitude of the
signal   was equal  to  one.  The frequencies   were  taken at   random   in the  interval
$[-B/2,B/2]$, where $B$ is the bandwidth assigned to the signal.
 
\item Sum of 200 delayed sinc pulses with random delays and amplitudes
\be{eq:320}
\sum_{p=1}^{200} a_p \mathrm{sinc}(B(t-\tau_p)).
\ee
The delays were  taken  at random in   $R(0,2T)$, knowing that   the sampling interval  is
$R(0,T)$. The amplitudes $a_p$  had random absolute  value in $[0,1]$  and random phase in
$[0,2\pi]$. They were then scaled so that the peak amplitude of the signal was one.

\end{enumerate}
In this example the time and frequency  variables were normalized so  that $T=1$.  For the
five components,  the  component index  ($m$), type  of  signal (type),  central frequency
($f_c$), bandwidth ($B$), bandwidth after windowing ($B'$), and the first ($f_a$) and last
($f_b$) frequencies in the trigonometric approximation were the following:
\begin{center}
\begin{tabular}{c|c|c|c|c|c|c}
$m$ & type & $f_c$  &   $B$   &  $B'$   & $f_a$ &  $f_b$  \\\hline
1   & 1    & 308.892 & 60.4428 & 69.5628 &  275  &  343\\
2   & 2    & 596.276 & 41.7585 & 50.8785 &  571  &  621\\
3   & 3    & 920.824 & 39.9765 & 49.0965 &  897  &  945\\
4   & 1    & 1169.11 & 66.6665 & 75.7865 &  1132 &  1207\\
5   & 2    & 1381.22 & 19.1557 & 28.2757 &  1368 &  1395\\
\end{tabular}
\end{center}
These bands are schematically plotted in Fig. \ref{fig:3}.
A set of moduli $Q$ was selected using the method in Sec. \ref{sec:lss}, 
[around Eq. (\ref{eq:57})]. The set was
\be{eq:97}
Q_k = 67+k,\; k=1,\ldots,\; K,\;\eqwith K=4.
\ee
This set  ensured that the  linear system in  (\ref{eq:57}) had full  column  rank and was
minimally  over-determined.   Actually, the  number of  samples was  above  the  number of
unknown coefficients just by one; (274 samples  were taken, but  there were 273 unknowns).
However, the noise factor in (\ref{eq:56}) was $\Fgam(I_{zw})=48.75$  dB, that is, there
were interpolation instants in which the signal-to-noise ratio would be degraded 48.75 dB,
(assuming  complex white noise). In  order to reduce this  figure,  several sampling grids
were added.  The final set of $Q$ values was
\be{eq:86}
11, 18, 19, 37, 49, 68, 69, 70, 71.
\ee
These  additional $Q$ values  were selected so   as to reduce   $\Fgam(I_{zw})$ as much as
possible   sequentially.  With   the  set   in  (\ref{eq:86}),   the   noise figure   was
$\Fgam(I_{zw})=18.77$ dB, i.e,  in the worst  case the noise  would be amplified 18.77 dB.
Fig.  \ref{fig:4} shows $\Fgam(t,I_{zw})$. 
Note that except  for  a few  peaks,  $\Fgam(t,I_{zw})$ is well   below this figure.   The
performance in terms of bandwidths and rates was the following:

\vspace{0.2cm}
\begin{tabular}{rl}
Nyquist bandwidth: & 1112.13\\
Landau limit: & 228\\
Landau limit after windowing: & 273.6\\
Sampling rate:& 394\\
(Nyquist rate)/(Sampling rate): &2.82\\
(Sampling rate)/(Landau limit): &1.73 
\end{tabular}
\vspace{0.2cm}

\noindent  The sampling rate  equaled 394, that  is, it was only   necessary to take this
number of samples in each period $T$. The  ratio between the sampling  rate and the Landau
limit shows how far this implementation was from the minimal  sampling rate (factor 1.73).
This  extra over-sampling was produced  by the windowing  and by the over-determination of
the linear system  so as to  reduce $\Fgam(I_{zw})$.  Relative to  the usual Nyquist rate,
this implementation afforded a sampling rate reduction of factor 2.82.

As commented in Sec.  \ref{sec:lss}, the  linear system that must  be solved for obtaining
the coefficients $\delta_p$ is that in  (\ref{eq:57}), which is a  sparse system.  In this
example,  the matrix corresponding  to (\ref{eq:57}) has only  2.18\% of non-zero elements
which are equal to one.   Thus, solving this  system using an  efficient method for sparse
linear systems,  like  that in  \cite{Paige82},  could  be  more  efficient than  directly
multiplying by the pseudo-inverse of the matrix corresponding to (\ref{eq:57}).

\begin{figure}
\centerline{\includegraphics{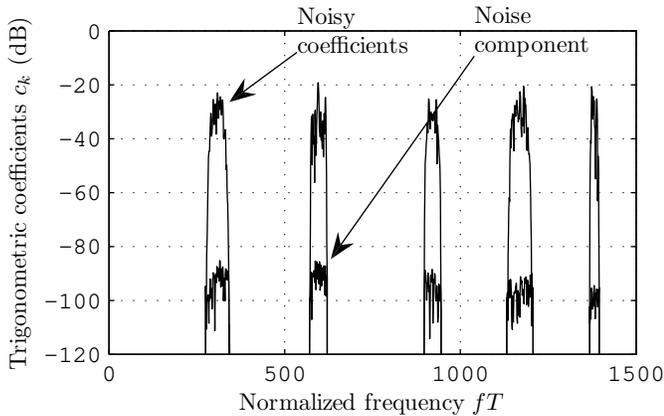}}
\caption{\label{fig:5} Noisy trigonometric coefficients $c_k$ and their noise component in
Sec. \ref{sec:ppn}.}
\end{figure}

Fig.   \ref{fig:5}  shows the  module  of the  coefficients  $\delta_p$  in  this example,
assuming that the ratio between the power of the multiband signal and the noise was 70 dB.
The noise component in these  coefficients is also  plotted in this  figure. Note that the
coefficients $\delta_p$ are correctly  estimated with signal-to-noise  ratio equal to 54.6
dB.

\begin{figure}
\centerline{\includegraphics{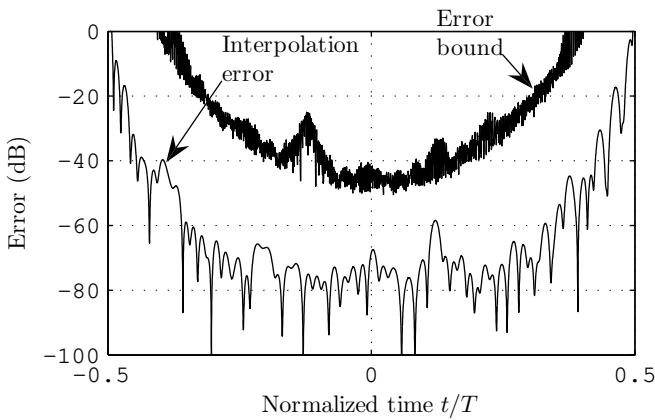}}
\caption{\label{fig:6} Interpolation error for the first multiband component in example in
  Sec \ref{sec:ppn}, and corresponding error bound in (\ref{eq:80}).}
\end{figure}

Finally, Fig. \ref{fig:6}  shows the error  in the  interpolation  of the first  multiband
component, together with the error bound in (\ref{eq:80}). Note that the error is small in
a wide range and that the bound is very conservative.

\subsection{Blind sampling}
\label{sec:bs}

\begin{figure}
\subfigure[$P=|I_{zw}|$]{\label{fig:7}\includegraphics{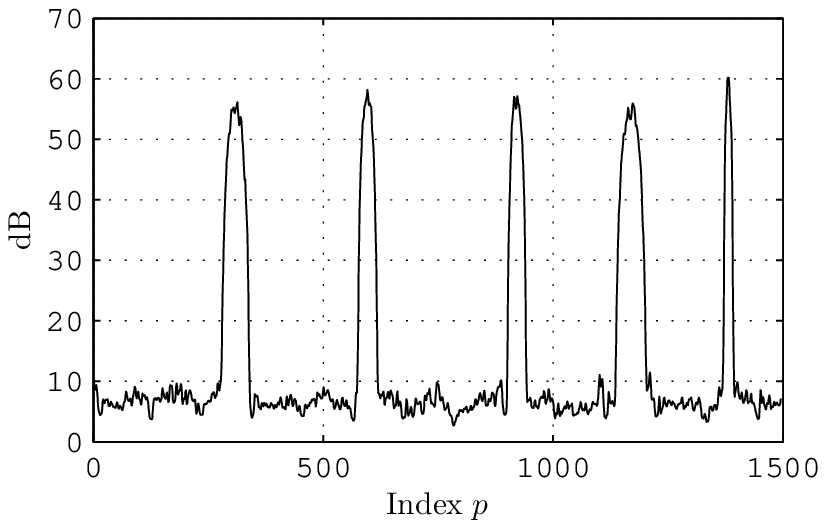}}
\subfigure[$P=|I_{zw}|-50$.]{\label{fig:8}\includegraphics{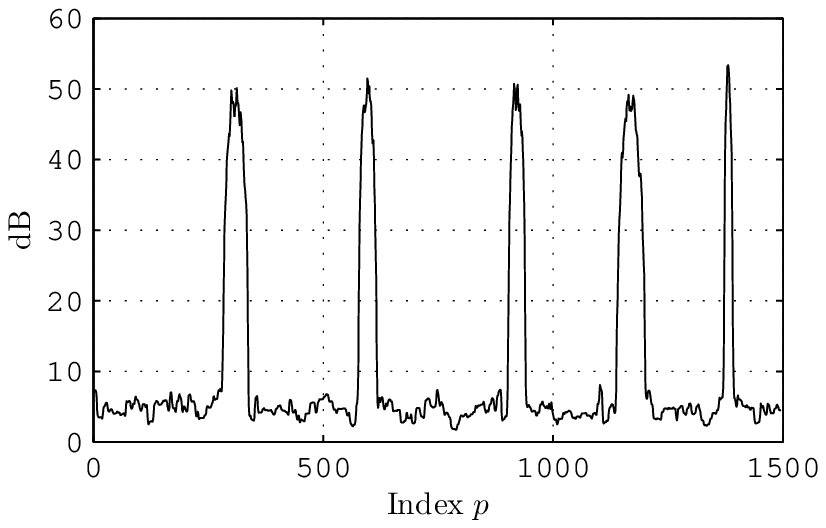}}
\subfigure[$P=|I_{zw}|+50$]{\label{fig:9}\includegraphics{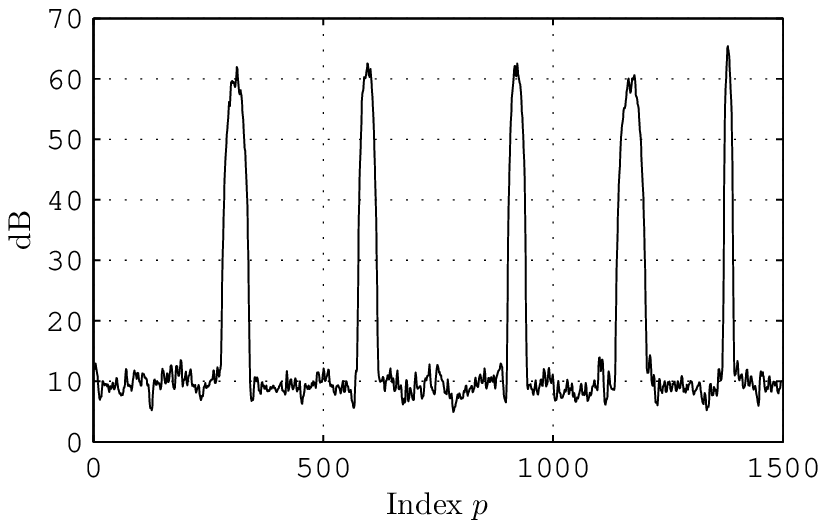}}
\caption{MUSIC spectra in Sec. \ref{sec:bs}.}
\end{figure}

In order to test the MUSIC algorithm in
Sec. \ref{sec:bsc}, the example in the previous section was repeated but for QPSK signals
(in order to reduce the computational load). Also, the set of moduli $Q_k$ was changed to 
\be{eq:98}
20, 50, 70, 110, 130,
\ee
so that all moduli $Q_k$ have common factor 10. Due to this  property, the sampling scheme
had period $T/10$, which  made it possible  to use the  central instants $\tau_h=T  h/10$,
i.e, the multiband signal was windowed in $H=500$ consecutive intervals $R(T h/10,T)$ with
90\% overlap.  Fig.  \ref{fig:7} shows the MUSIC spectrum in Eq. (\ref{eq:95}) assuming a
signal subspace of dimension $P=|I_{zw}|$.  Note that this estimator is able to detect the
multiband components as well as their  approximate width.  (Compare  this figure with Fig.
\ref{fig:3}.)  Figs.  \ref{fig:8}  and \ref{fig:9} show the  same spectrum but assuming  a
signal subspace of dimension $|I_{zw}|-50$ and  $|I_{zw}|+50$, respectively. Note that the
detection  capability remains  unchanged,   relative to the   case  $P=|I_{zw}|$ in   Fig.
\ref{fig:7}.

\section{Conclusions}

A method for sampling bounded multiband signals has been presented, that makes it possible
to approach the Landau limit, while  keeping at the  same time the  noise sensitivity at a
low level.  The method is based on approximating  the product of  the signal with a window
function   by  means of   a  trigonometric polynomial.  This   polynomial ``inherits'' the
multiband  property  from the  signal.  The  associated sampling  scheme   is the recently
proposed synchronous multi-rate sampling.  Also, it was shown that the blind sampling of a
bounded  multiband signal can  be reduced to  the  sampling of  a  collection of multiband
trigonometric polynomials with unknown index support.  This fact made it possible to apply
a DOA  estimation  algorithm (MUSIC)  for  the  detection of  the  spectral  support.  The
performance of the methods in the paper have been validated in several numerical examples.
 
\section{Acknowledgments}

The author would like to thank the editor and reviewers of this paper for their thorough
revision and the many useful comments.  

\appendices

\section{Bound on the tails of the window function}
\label{ap:btw}

The computation  of the error  bound $\epsilon$ in Eq.  (\ref{eq:18}) depends on bounding
the series
\be{eq:99}
\FS(a,b)\equiv\sum_{n=0}^\infty \frac{1}{(a+n)\sqrt{(a+n)^2-b^2}}
\ee
for constants $a>b>0$. For this, introduce first the variable change
\be{eq:100}
x(n)=\Big(\frac{b}{a+n}\Big)^2
\ee
to obtain
\bae{eq:101}{1.5}{r@{\,=\,}l}
\D{\FS(a,b)}&\D{\sum_{n=0}^\infty \frac{1}{(a+n)^2}\frac{1}{\sqrt{1-(b/(a+n))^2}}
}\\
 &\D{
 \frac{1}{b^2}\sum_{n=0}^\infty \frac{x(n)}{\sqrt{1-x(n)}}.
 }
\eae
The  function $1/\sqrt{1-x}$ is  increasing in $[0,1[$  and  the maximum  value  of $x=x(n)$ in
(\ref{eq:100}) is $(b/a)^2$. So, this function can be bounded as
\be{eq:102}
\frac{1}{\sqrt{1-x(n)}}\leq \frac{1}{\sqrt{1-(b/a)^2}}=\frac{a}{\sqrt{a^2-b^2}}.
\ee
Coming back to (\ref{eq:101}), it follows that
\bae{eq:103}{1.5}{r@{\,}c@{\,}l}
\D{\FS(a,b)}&\D{\leq}&\D{\frac{a}{b^2\sqrt{a^2-b^2}}\sum_{n=0}^\infty x(n)=
\frac{a}{\sqrt{a^2-b^2}}\sum_{n=0}^\infty \frac{1}{(a+n)^2}} \\
 &\D{=}&\D{\frac{1}{\sqrt{1-(b/a)^2}}
\Fpsi'(a),}
\\
\eae
where $\Fpsi'(a)$ is the derivative of the polygamma function. 

Let us proceed   with the derivation  of  the bound. Recalling   the definitions  in  Eqs.
(\ref{eq:38}) and (\ref{eq:39}), since the sine function is bounded by one, it is
\bae{eq:104}{1.5}{l}
\D{|\Fw(t)|\leq \frac{1}{\pi\delta B_w t}\frac{1}{\pi (1-\delta)B_w \sqrt{t^2-\rho^2 T^2/4}}
\hspace{0.5cm}{}}\\
\D{{}\hfill \cdot\frac{1}{\mathrm{sinc}(j(1-\delta)\rho B_wT/2)}}\\
\D{=\frac{1}{\mathrm{sinc}(j(1-\delta)\rho B_wT/2)\pi^2\delta(1-\delta)
B_w^2 t \sqrt{t^2-\rho^2 T^2/4}}}\\
\D{=C\frac{1}{t \sqrt{t^2-\rho^2 T^2/4}},}
\eae
where $C$ is the constant
\be{eq:105}
C\equiv\big(\mathrm{sinc}(j(1-\delta)\rho B_wT/2)
\pi^2\delta(1-\delta)B_w^2\big)^{-1}. 
\ee
Now,
\bae{eq:106}{1.5}{l}
\D{\sum_{p=2}^\infty |\Fw(t+pT)|\leq C \sum_{p=2}^\infty 
\frac{1}{(t+pT) \sqrt{(t+pT)^2-\rho^2 T^2/4}}}\\
\D{=\frac{C}{T^2}\sum_{p=2}^\infty \frac{1}{(t/T+p) \sqrt{(t/T+p)^2-\rho^2/4}} }\\
\D{=\frac{C}{T^2}\FS(2+t/T,\rho/2).}
\eae
Also,
\bae{eq:107}{1.5}{l}
\D{\sum_{p=2}^\infty |\Fw(t-pT)|=\sum_{p=2}^\infty |\Fw(-t+pT)|\leq 
\frac{C}{T^2}\FS(2-t/T,\rho/2).}
\eae
So, combining the last two inequalities, it is 
\bae{eq:108}{1.5}{l}
\D{\sum_{|p|> 1}|\Fw(t+pT)|\hspace{1cm}{}}\\
\D{{}\hfill\leq \frac{C}{T^2}(\FS(2-t/T,\rho/2)+\FS(2+t/T,\rho/2)),}
\eae
and
\bae{eq:109}{1.5}{l}
\D{\sum_{p\neq 0}|\Fw(t+pT)|\leq|\Fw(t-T)|+|\Fw(t+T)|}\\
\D{\hfill + \frac{C}{T^2}(\FS(2-t/T,\rho/2)+\FS(2+t/T,\rho/2)),}
\eae
The final bound is obtained by substituting (\ref{eq:103}) and (\ref{eq:105}) into this inequality.

\begin{table}[h]
\begin{tabular}{rl}
$\tilde{\;}$& Periodic repetition with period $T$, Eq. (\ref{eq:5})\\
$|\cdot|$& \parbox[t]{6.8cm}{$|J|$ is the number of elements of the finite set $J$}\\
$\ma$& Data matrix in (\ref{eq:93})\\
$[a_m,b_m]$& Spectral support of $\Fs_m(t)$\\
$A_\eta$& $l^2$ norm of set of samples $\Feta(t_n)$ in (\ref{eq:75})\\
$A_s$ & Amplitude of $\Fs(t)$, $|\Fs(t)|\leq A_s$\\
$A_{s,m}$ & Amplitude of $\Fs_m(t)$\\
$\hat{\alpha}_{h,n}$& Data value in (\ref{eq:91})\\
$\Falp(t_n;J)$ &\parbox[t]{6.8cm}{Generic sparse trigonometric polynomial in
  (\ref{eq:20})}\\
$\Falp(t_n;J)$ & Trigonometric polynomial in Eq. (\ref{eq:90})\\
$B$& Two-sided bandwidth of $\Fs(t)$\\
$B_w$& Two-sided bandwidth of $\Fw(t)$, Sec. \ref{sec:spm}\\
$\mathcal{B}_{\pi  B}^\infty$&Bernstein space of type $\pi B$\\
$\beta_p$& Coefficient of $\Falp(t;J)$ in (\ref{eq:20}) \\
$\beta_{h,p}$& Coefficient of $\Falp_h(t;J)$ in (\ref{eq:90}) \\
$\mb$ & Coefficient matrix in (\ref{eq:93})\\
$\Fc_p(\tau)$& Coefficient of $\tilde{\Fz}_w(t;\tau)$\\
$\delta$ & Window parameter in Eq. (\ref{eq:38})\\
$\delta_p$& Unknowns in sparse linear system, Eq. (\ref{eq:43})\\
$\delta_w$& Lower bound on $\Fw(t)$ in $R(0,T_1)$, $\delta_w>0$, Eq. (\ref{eq:41})\\
$\eta_{p,n}$& Pseudo-inverse of the linear system in (\ref{eq:21})\\
$\vphi(p)$& Vector of exponentials in (\ref{eq:93})\\
$\mphi$& Matrix of exponentials in (\ref{eq:93})\\
$\Fgam(J')$ & Supremum of sensitivity measure in $R(0,T)$, Eq. (\ref{eq:56}) \\
$\Fgam(t,J',t_0)$ & Sensitivity measure in Eq. (\ref{eq:55}) or (\ref{eq:76})\\
$\Fh_{m,n}(t)$& Generic interpolation functions in (\ref{eq:9})\\
$I_{sw,m}$& Index set of polynomial $\tilde{\Fs}_{w,m}(t;\tau)$, Eq. (\ref{eq:14})\\
$I_{zw}$& Index set of polynomial $\tilde{\Fz}_w(t;\tau)$, Eq. (\ref{eq:15}) \\
$\mi$& Identity matrix in Sec. \ref{sec:bsc}\\
$J$, $J'$& Arbitrary finite sets of integers (indices)\\
$\lambda_{p,k,r}$& Pseudo-inverse of linear system in (\ref{eq:49})\\
$M$&Number of multiband components\\
$\mn$& Noise matrix in (\ref{eq:93})\\ 
$n_{k,q}(t_0)$& Integer shift in Eq. (\ref{eq:61})\\
$p(r)$ & Ordering of set $J$ in (\ref{eq:92})\\
$\vpr$ & Vector of indices in (\ref{eq:93})\\
$R(\tau,T)$& Time interval defined in (\ref{eq:4})\\
$\rho$ & Window parameter in (\ref{eq:39})\\
$\Fs(t)$ & Generic band-limited signal\\
$\Fs_m(t)$ &Component of $\Fz(t)$\\
$\Fs_{w,m}(t;\tau)$ & Windowed multiband component in (\ref{eq:31})\\
$\tilde{\Fs}_{w,m}(t;\tau)$ & Periodized version of $\Fs_{w,m}(t;\tau)$, 
Eq. (\ref{eq:34}) \\
$\mathcal{S}_z$&Spectral support of $\Fz(t)$ in (\ref{eq:8})\\
$\mathcal{S}_{zw}$& \parbox[t]{6.8cm}{Spectral support of 
$\tilde{\Fz}_w(t;\tau)=\Fz(\tau+t)\Fw(t)$ in Eq. (\ref{eq:13})}\\
$t_n$& Generic sampling instants in $R(0,T)$\\
$T$& Length of the interpolation interval, Sec. \ref{sec:spm} \\
$T_1$&\parbox[t]{6.8cm}{Length of interval $R(\tau,T_1)$ in which the interpolation is 
accurate, Sec. \ref{sec:spm}}\\
$\tau$&\parbox[t]{6.8cm}{Instant around which $\Fz(t)$ is interpolated,
  Sec. \ref{sec:spm}}\\
$\mur$& Matrix spanning the noise subspace of $\ma$, Sec. \ref{sec:bsc}\\
$\mu_{h,n}$& Noise sample in model (\ref{eq:91})\\ 
$\Fthe_n(t;J)$ & Interpolation polynomials in (\ref{eq:23})\\
$\Fthe_n(t;J,t_0)$ & Interpolation polynomials in (\ref{eq:53})\\
$\Fw(t)$& Band-limited window, Sec. \ref{sec:spm}\\
$\Fz(t)$& Multiband signal with bounded components\\
$\Fz_w(t;\tau)$ & Windowed multiband signal in (\ref{eq:29})\\ 
$\tilde{\Fz}_w(t;\tau)$& Periodized version of $\Fz_w(t;\tau)$, Eq. (\ref{eq:35})\\
\end{tabular}
\caption{\label{tab:1} List of symbols}
\end{table}

\bibliographystyle{IEEEbib}

\bibliography{../../../Utilities/LaTeX/Bibliography}

\end{document}

%% file: MatrixNotation.tex
\newlength{\Lpr}
\newsavebox{\Bpr}

\newcommand{\D}[1]{\ensuremath{\displaystyle #1}}

\newcommand{\V}[1]{\mbox{\boldmath$\mathbf{#1}$\unboldmath}}

\newcommand{\Herm}{\ensuremath{^{\mathrm{H}}}}

\newcommand{\sinc}{\ensuremath{{\mathrm{sinc}}}}

\newcommand{\bdm}{\begin{displaymath}}
\newcommand{\edm}{\end{displaymath}}

\newcommand{\be}[1]{\begin{equation} \label{#1}}
\newcommand{\ee}{\end{equation}}

\newcommand{\bae}[3]{
\begin{equation} \label{#1}
\renewcommand{\arraystretch}{#2}
\begin{array}{#3}}

\newcommand{\eae}{\end{array}\end{equation}}

\newcommand{\baen}[2]{
\begin{displaymath} 
\renewcommand{\arraystretch}{#1}
\begin{array}{#2}}

\newcommand{\eaen}{\end{array}\end{displaymath}}

\newcommand{\DefLetter}[4]{
\newcommand{#1}{\ensuremath{\V{#2}}} 
\newcommand{#3}{\ensuremath{\V{#4}}} 
}

\DefLetter{\vzer}{0}{\mzer}{0}
\DefLetter{\vone}{1}{\mone}{1}
\DefLetter{\va}{a}{\ma}{A}
\DefLetter{\vb}{b}{\mb}{B}
\DefLetter{\vc}{c}{\mc}{C}
\DefLetter{\vd}{d}{\md}{D}
\DefLetter{\ve}{e}{\me}{E}
\DefLetter{\vf}{f}{\mf}{F}
\DefLetter{\vg}{g}{\mg}{G}
\DefLetter{\vh}{h}{\mh}{H}
\DefLetter{\vi}{i}{\mi}{I}
\DefLetter{\vj}{j}{\mj}{J}
\DefLetter{\vk}{k}{\mk}{K}
\DefLetter{\vl}{l}{\ml}{L}
\DefLetter{\vm}{m}{\mm}{M}
\DefLetter{\vn}{n}{\mn}{N}
\DefLetter{\vpr}{p}{\mpr}{P}
\DefLetter{\vq}{q}{\mq}{Q}
\DefLetter{\vr}{r}{\mr}{R}
\DefLetter{\vs}{s}{\ms}{S}
\DefLetter{\vt}{t}{\mt}{T}
\DefLetter{\vur}{u}{\mur}{U}
\DefLetter{\vv}{v}{\mv}{V}
\DefLetter{\vw}{w}{\mw}{W}
\DefLetter{\vx}{x}{\mx}{X}
\DefLetter{\vy}{y}{\my}{Y}
\DefLetter{\vz}{z}{\mz}{Z}

\DefLetter{\vdel}{\delta}{\mdel}{\Delta}
\DefLetter{\vphi}{\phi}{\mphi}{\Phi}
\DefLetter{\vpsi}{\psi}{\mpsi}{\Psi}
\DefLetter{\vrho}{\rho}{\mrho}{\Lambda}
\DefLetter{\vxi}{\xi}{\mxi}{\Xi}

\DefLetter{\valpha}{\alpha}{\malpha}{\Alpha}
\DefLetter{\vbeta}{\beta}{\mbeta}{\Beta}
\DefLetter{\vlam}{\lambda}{\mlam}{\Lambda}
\DefLetter{\vsig}{\sigma}{\msig}{\Sigma}
\DefLetter{\vtau}{\tau}{\mtau}{\tau}
\DefLetter{\vtheta}{\theta}{\mtheta}{\Theta}
\DefLetter{\vome}{\omega}{\mome}{\Omega}
\DefLetter{\vzero}{0}{\mzero}{0}
\DefLetter{\vgam}{\gamma}{\mgam}{\Gamma}
\DefLetter{\veps}{\epsilon}{\meps}{\Epsilon}
\DefLetter{\veta}{\eta}{\meta}{\Eta}

\newcommand{\eqwith}{\:\:\mbox{with}\:\:}

\newcommand{\DefFuncLetter}[2]{
\newcommand{#1}{\ensuremath{{\mathrm{#2}}}} 
}

\DefFuncLetter{\Fzer}{0}
\DefFuncLetter{\Fa}{a}
\DefFuncLetter{\FA}{A}
\DefFuncLetter{\Fb}{b}
\DefFuncLetter{\Fc}{c}
\DefFuncLetter{\FC}{C}
\DefFuncLetter{\Fd}{d}
\DefFuncLetter{\Fe}{e}
\DefFuncLetter{\Ff}{f}
\DefFuncLetter{\Fg}{g}
\DefFuncLetter{\FG}{G}
\DefFuncLetter{\Fh}{h}
\DefFuncLetter{\FH}{H}
\DefFuncLetter{\Fi}{i}
\DefFuncLetter{\Fk}{k}
\DefFuncLetter{\Fl}{l}
\DefFuncLetter{\FL}{L}
\DefFuncLetter{\Fm}{m}
\DefFuncLetter{\Fn}{n}
\DefFuncLetter{\Fnr}{n}
\DefFuncLetter{\FN}{N}
\DefFuncLetter{\Fo}{o}
\DefFuncLetter{\FO}{O}
\DefFuncLetter{\Fpr}{p}
\DefFuncLetter{\FPr}{P}
\DefFuncLetter{\Fq}{q}
\DefFuncLetter{\Fr}{r}
\DefFuncLetter{\Fs}{s}
\DefFuncLetter{\FS}{S}
\DefFuncLetter{\Ft}{t}
\DefFuncLetter{\FT}{T}
\DefFuncLetter{\Fu}{u}
\DefFuncLetter{\FU}{U}
\DefFuncLetter{\Fv}{v}
\DefFuncLetter{\Fw}{w}
\DefFuncLetter{\FW}{W}
\DefFuncLetter{\Fx}{x}
\DefFuncLetter{\Fy}{y}
\DefFuncLetter{\FY}{Y}
\DefFuncLetter{\Fz}{z}
\DefFuncLetter{\FZ}{Z}
\DefFuncLetter{\Falp}{\alpha}
\DefFuncLetter{\Fbet}{\beta}
\DefFuncLetter{\Fchi}{\chi}
\DefFuncLetter{\Fdel}{\delta}
\DefFuncLetter{\Fzet}{\zeta}
\DefFuncLetter{\Feta}{\eta}
\DefFuncLetter{\Fphi}{\phi}
\DefFuncLetter{\FPhi}{\Phi}
\DefFuncLetter{\Fpsi}{\psi}
\DefFuncLetter{\FPsi}{\Psi}
\DefFuncLetter{\Fgam}{\gamma}
\DefFuncLetter{\FGam}{\Gamma}
\DefFuncLetter{\Flam}{\lambda}
\DefFuncLetter{\Fsig}{\sigma}
\DefFuncLetter{\Ftau}{\tau}
\DefFuncLetter{\Fome}{\omega}
\DefFuncLetter{\Feps}{\epsilon}
\DefFuncLetter{\Fthe}{\theta}
\DefFuncLetter{\Fvar}{\vartheta}

\DefFuncLetter{\FB}{B}
\DefFuncLetter{\FD}{D}
\DefFuncLetter{\FE}{E}
\DefFuncLetter{\FF}{F}
\DefFuncLetter{\FI}{I}
\DefFuncLetter{\FJ}{J}
\DefFuncLetter{\FM}{M}
\DefFuncLetter{\FR}{R}
\DefFuncLetter{\FX}{X}

\newcommand{\DefCalLetter}[2]{
\newcommand{#1}{\ensuremath{\mathcal{#2}}} 
}
\DefCalLetter{\CC}{C}
\DefCalLetter{\CD}{D}

\DefCalLetter{\CS}{S}
\DefCalLetter{\CV}{V}

\newcommand{\DefSubLetter}[2]{
\newcommand{#1}{\mathrm{#2}} 
}

\DefSubLetter{\slzer}{0}
\DefSubLetter{\sla}{a}
\DefSubLetter{\slA}{A}
\DefSubLetter{\slb}{b}
\DefSubLetter{\slB}{B}
\DefSubLetter{\slc}{c}
\DefSubLetter{\slC}{C}
\DefSubLetter{\sld}{d}
\DefSubLetter{\slD}{D}
\DefSubLetter{\sle}{e}
\DefSubLetter{\slE}{E}
\DefSubLetter{\slf}{f}
\DefSubLetter{\slF}{F}
\DefSubLetter{\slg}{g}
\DefSubLetter{\slG}{G}
\DefSubLetter{\slh}{h}
\DefSubLetter{\slH}{H}
\DefSubLetter{\sli}{i}
\DefSubLetter{\slI}{I}
\DefSubLetter{\slk}{k}
\DefSubLetter{\sll}{l}
\DefSubLetter{\slL}{L}
\DefSubLetter{\slm}{m}
\DefSubLetter{\slM}{M}
\DefSubLetter{\sln}{n}
\DefSubLetter{\slnr}{n}
\DefSubLetter{\slN}{N}
\DefSubLetter{\slo}{o}
\DefSubLetter{\slp}{p}
\DefSubLetter{\slP}{P}
\DefSubLetter{\slq}{q}
\DefSubLetter{\slQ}{Q}
\DefSubLetter{\slr}{r}
\DefSubLetter{\slR}{R}
\DefSubLetter{\sls}{s}
\DefSubLetter{\slS}{S}
\DefSubLetter{\slt}{t}
\DefSubLetter{\slT}{T}
\DefSubLetter{\slu}{u}
\DefSubLetter{\slU}{U}
\DefSubLetter{\slv}{v}
\DefSubLetter{\slw}{w}
\DefSubLetter{\slW}{W}
\DefSubLetter{\slx}{x}
\DefSubLetter{\slX}{X}
\DefSubLetter{\sly}{y}
\DefSubLetter{\slY}{Y}
\DefSubLetter{\slz}{z}
\DefSubLetter{\slZ}{Z}

\DefSubLetter{\slalp}{\alpha}
\DefSubLetter{\slbet}{\beta}
\DefSubLetter{\sldel}{\delta}
\DefSubLetter{\slDel}{\Delta}
\DefSubLetter{\sleps}{\epsilon}
\DefSubLetter{\slgam}{\gamma}
\DefSubLetter{\slphi}{\phi}
\DefSubLetter{\sltau}{\tau}
\DefSubLetter{\slxi}{\xi}
\DefSubLetter{\slthe}{\theta}


%% file: Trans2010.bbl
\begin{thebibliography}{10}

\bibitem{Jerri77}
Abdul~J. Jerri,
\newblock ``The {Shannon} {Sampling} {Theorem} -- its various extensions and
  applications: A tutorial review,''
\newblock {\em Proceedings of the IEEE}, vol. 65, no. 11, pp. 1565--1596, Nov
  1977.

\bibitem{Landau67}
H.~J. Landau,
\newblock ``Sampling, data transmission, and the {N}yquist rate,''
\newblock {\em Proceedings of the IEEE}, vol. 55, no. 10, pp. 1701--1706, Oct
  1967.

\bibitem{Herley99}
C.~Herley and Ping~Wah Wong,
\newblock ``Minimum rate sampling and reconstruction of signals with arbitrary
  frequency support,''
\newblock {\em IEEE Transactions on Information Theory}, vol. 45, no. 5, pp.
  1555--1564, July 1999.

\bibitem{Feng96}
P.~Feng, S.~F. Yau, and Y~Bresler,
\newblock ``A multicoset sampling approach to the missing cone problem in
  computer-aided tomography,''
\newblock in {\em IEEE International Symposium on Circuits and Systems, ISCAS
  '96}, May 1996, vol.~2, pp. 734--737.

\bibitem{Feng96b}
P.~Feng and Y.~Bresler,
\newblock ``Spectrum-blind minimum-rate sampling and reconstruction of
  multiband signals,''
\newblock in {\em IEEE International Conference on Acoustics, Speech, and
  Signal Processing}, May 1996, vol.~3, pp. 1688 --1691.

\bibitem{Venkataramani00}
R.~Venkataramani and Y.~Bresler,
\newblock ``Perfect reconstruction formulas and bounds on aliasing error in
  sub-{N}yquist nonuniform sampling of multiband signals,''
\newblock {\em IEEE Transactions on Information Theory}, vol. 46, no. 6, pp.
  2173--2183, Sept 2000.

\bibitem{Venkataramani01}
R.~Venkataramani and Y.~Bresler,
\newblock ``Optimal sub-{N}yquist nonuniform sampling and reconstruction for
  multiband signals,''
\newblock {\em IEEE Transactions on Signal Processing}, vol. 49, no. 10, pp.
  2301--2313, Oct 2001.

\bibitem{Mishali08}
M.~Mishali and Y.~C. Eldar,
\newblock ``Spectrum-blind reconstruction of multi-band signals,''
\newblock in {\em Proc. IEEE International Conference on Acoustics, Speech and
  Signal Processing ICASSP 2008}, Mar. 2008, pp. 3365--3368.

\bibitem{Mishali09}
Y.~C.~Eldar M.~Mishali,
\newblock ``Blind multiband signal reconstruction: compressed sensing of analog
  signals,''
\newblock {\em IEEE Transactions on Signal Processing}, vol. 57, no. 3, pp.
  993--1009, Mar 2009.

\bibitem{Tropp10}
J.~A. Tropp, Laska~J. N., M.~F. Duarte, J.~K. Romberg, and R.~G. Baraniuk,
\newblock ``Beyond nyquist: efficient sampling of sparse bandlimited signals,''
\newblock {\em IEEE Transactions on Information Theory}, vol. 56, no. 1, pp.
  520--544, Jan 2010.

\bibitem{Mishali10}
M.~Mishali and Y.~C. Eldar,
\newblock ``From theory to practice: Sub-{N}yquist sampling ofsparse wideband
  analog signals,''
\newblock arXiv.org 0902.4291; to appear in IEEE J. Sel. Topics Signal
  Processing.

\bibitem{Rosenthal08}
A.~Rosenthal, A.~Linden, and M.~Horowitz,
\newblock ``Multirate asynchronous sampling of sparse multiband signals,''
\newblock {\em J. Opt. Soc. Am.}, vol. 25, no. 9, pp. 2320--2330, Sept 2008.

\bibitem{Fleyer10}
Michael Fleyer, Alex Linden, Moshe Horowitz, and Amir Rosenthal,
\newblock ``Multirate synchronous sampling of sparse multiband signals,''
\newblock To appear in the IEEE Transactions on Signal Processing.

\bibitem{Butzer82}
Paul~L. Butzer, Wolfgang Engels, and Ursula Scheben,
\newblock ``Magnitude of the truncation error in sampling expansions,''
\newblock {\em IEEE Transactions on Acoustics, Speech, and Signal Processing},
  vol. ASSP-30, no. 6, pp. 906--912, Dec 1982.

\bibitem{Laakso96}
T.~I. Laakso, V.~V\"alim\"aki, M.~Karjalainen, and U.~K. Laine,
\newblock ``Splitting the {U}nit {D}elay,''
\newblock {\em IEEE Signal Processing Magazine}, vol. 13, no. 1, pp. 30--60,
  Jan 1996.

\bibitem{Higgins96}
J.~R. Higgins,
\newblock {\em Sampling Theory in {F}ourier and signal analysis.
  {F}oundations.},
\newblock Oxford Science Publications, 1996.

\bibitem{Chen10}
Y.~Chen, M.~Mishali, Y.~C. Eldar, and Alfred~O. Hero~III,
\newblock ``Modulated wideband converter with non-ideal lowpass filters,''
\newblock Proc. IEEE Int. Conf. Acoustics, Speech and Signal Processing
  (ICASSP-10).

\bibitem{Helms62}
H.~D. Helms,
\newblock ``Truncation error of sampling theorem expansions,''
\newblock {\em Proceedins of the {I}{R}{E}}, pp. 179--184, Feb 1962.

\bibitem{Selva06}
J.~Selva,
\newblock ``Interpolation of bounded band-limited signals and applications,''
\newblock {\em IEEE Transactions on Signal Processing}, vol. 54, no. 11, pp.
  4244--4260, Nov 2006.

\bibitem{Knab79}
J.~J. Knab,
\newblock ``Interpolation of band-limited functions using the {A}pproximate
  {P}rolate series,''
\newblock {\em IEEE Transactions on Information Theory}, vol. IT-25, no. 6, pp.
  717--720, Nov 1979.

\bibitem{Selva08}
J.~Selva,
\newblock ``An efficient structure for the design of {Variable} {Fractional}
  {Delay} filters based on the windowing method,''
\newblock {\em IEEE Transactions on Signal Processing}, vol. 56, no. 8, pp.
  3770--3775, Aug 2008.

\bibitem{Selva09}
J.~Selva,
\newblock ``Functionally weighted {L}agrange interpolation of band-limited
  signals from nonuniform samples,''
\newblock {\em IEEE Transactions on Signal Processing}, vol. 57, no. 1, pp.
  168--181, Jan 2009.

\bibitem{Selva09b}
J.~Selva,
\newblock ``Optimal variable fractional delay filters in time-domain
  {L}-infinity norm,''
\newblock in {\em International Conference on Acoustics Speech, and Signal
  Processing, ICASSP'-09}, Apr 2009, pp. 3373--3376.

\bibitem{Cotter05}
S.~F. Cotter, B.~D. Rao, K~Engan, and K.~Kreutz-Delgado,
\newblock ``Sparse solutions to linear inverse problems with multiple
  measurement vectors,''
\newblock {\em IEEE Transactions on Signal Processing}, vol. 53, no. 7, pp.
  2477--2488, Jul 2005.

\bibitem{VanTreesP4}
Harry~L. van Trees,
\newblock {\em Detection, Estimation, and Modulation Theory. Part IV},
\newblock John Wiley \& Sons, Inc, first edition, 2002.

\bibitem{Prasolov94}
V.~V. Prasolov,
\newblock {\em Problems and Theorems in Linear Algebra},
\newblock American Mathematical Society, 1994.

\bibitem{Schmidt86}
Ralph~O. Schmidt,
\newblock ``Multiple emitter location and signal parameter estimation,''
\newblock {\em IEEE Transactions on and Antennas Propagation}, vol. 34, no. 3,
  pp. 276--280, Mar. 1986.

\bibitem{Paige82}
C.~C. Paige and Saunders~M. A.,
\newblock ``{L}{S}{Q}{R}: Sparse equations and least squares,''
\newblock {\em ACM Transactions on Mathematical Software}, vol. 8, no. 1, pp.
  43--71, Mar 1982.

\end{thebibliography}
